\documentclass[amssymb,aps,prd,floatfix,twocolumn]{revtex4}
\usepackage[intlimits]{amsmath}
\usepackage{amssymb}
\usepackage{graphicx}
\usepackage{tensor}
\usepackage[dvipsnames]{xcolor}
\usepackage{hyperref}

\begin{document}

\title{Accretion of the relativistic Vlasov gas in the equatorial plane of the Kerr black hole}

\author{Adam Cie\'{s}lik}\email{adam.cieslik@doctoral.uj.edu.pl}
\affiliation{Instytut Fizyki Teoretycznej, Uniwersytet Jagiello\'{n}ski, {\L}ojasiewicza 11, 30-348 Krak\'{o}w, Poland}
\author{Patryk Mach}\email{patryk.mach@uj.edu.pl}
\affiliation{Instytut Fizyki Teoretycznej, Uniwersytet Jagiello\'{n}ski, {\L}ojasiewicza 11, 30-348 Krak\'{o}w, Poland}
\author{Andrzej Odrzywo{\l}ek}\email{andrzej.odrzywolek@uj.edu.pl}
\affiliation{Instytut Fizyki Teoretycznej, Uniwersytet Jagiello\'{n}ski, {\L}ojasiewicza 11, 30-348 Krak\'{o}w, Poland}

\begin{abstract}
    We investigate stationary accretion of the collisionless Vlasov gas onto the Kerr black hole, occurring in the equatorial plane. The solution is specified by imposing asymptotic boundary conditions: at infinity the gas obeys the Maxwell-J\"{u}ttner distribution, restricted to the equatorial plane (both in positions and momenta). In the vicinity of the black hole, the motion of the gas is governed by the spacetime geometry. We compute accretion rates of the rest-mass, the energy, and the angular momentum, as well as the particle number surface density, focusing on the dependence of these quantities on the asymptotic temperature of the gas and the black hole spin. The rest-mass and energy accretion rates, normalized by the black hole mass and appropriate asymptotic surface densities of the gas, increase with increasing asymptotic temperature. The accretion slows down the rotation of the black hole.    
\end{abstract}

\maketitle

\section{Introduction}

General-relativistic kinetic theory provides a description of gases in strong gravitational fields, alternative to the more popular hydrodynamic approach. When collisions between gas particles are rare and the mean free path is large, fluid dynamics does not apply. In this case, a more reasonable approximation is provided by the kinetic theory of non-colliding gas particles. A standard kinetic description uses a distribution function defined on a suitable one-particle phase space. In general-relativistic literature this approximation is sometimes referred to as the Vlasov gas. Vlasov models are often used in stellar dynamics to describe galaxies and clusters of galaxies (see a mathematical analysis in \cite{Rein1,Rein2} or \cite{BT} for the Newtonian account). Another physical realization of Vlasov systems would be in the form of non-interacting dark matter massive particles \cite{DM_Vlasov}. Classic reviews of the general-relativistic kinetic theory can be found e.g.\ in \cite{Synge,israel,Ehlers1,Ehlers2,Groot,Cercignani}; see also \cite{AndreassonReview,zannias,Acuna} for more recent accounts.

Rigorous relativistic models of accretion flows onto black holes derived within the framework of the kinetic theory are relatively recent. Spherically symmetric steady accretion of the collisionless Vlasov gas onto a Schwarzschild black hole (a kinetic counterpart of the Michel accretion model \cite{Michel,Chaverra}) has been investigated in 2017 by Rioseco and Sarbach \cite{Olivier1, Olivier2}. An extension of this model for Reissner-Nordstr\"{o}m black holes is given in \cite{cieslik}. Accretion of the collisionless Vlasov gas onto a moving Schwarzschild black hole (a general-relativistic counterpart of the Bondi-Hoyle-Lyttleton accretion problem) has been investigated in 2021 \cite{pmao1,pmao2,pmao3}. In the same year Gamboa et al.\ described accretion of the Vlasov gas onto a Schwarzschild black hole occurring from a sphere of a finite radius \cite{Gamboa}. A kinetic description of the phase-space mixing at the equatorial plane of the Kerr spacetime is given in \cite{Olivier3}. Some aspects of the dynamics of the massless Vlasov gas in a slowly rotating Kerr spacetime are analysed in \cite{Andersson}.  We would also like to mention recent studies of self-gravitating static Vlasov configurations around black holes \cite{Andreasson, Gunther}.

In this paper we derive stationary axially symmetric solutions of the collisionless Vlasov equation corresponding to a gas accreting onto a Kerr black hole at the equatorial plane. They constitute a model of an infinite, geometrically thin accretion disk. At infinity, the gas is assumed to be described by the Maxwell-J\"{u}ttner distribution confined to the equatorial plane. The flow of the gas in the vicinity of the black hole is still restricted to the equatorial plane, but otherwise it is governed by the geometry of the spacetime. In a sense, our model is an analogue of the spherically symmetric one obtained for Schwarzschild black holes in \cite{Olivier1}. The restriction to the equatorial plane of the Kerr black hole is justified by the algebraic simplicity of equations governing the geodesic motion at the equatorial plane. On the other hand, this model seems to be of interest on its own, as a model of a thin accretion disk around the Kerr black hole. It allows us to provide explicit expressions for the particle current surface density, as well as mass and angular momentum accretion rates. We find for instance that the accretion process described by this model slows down the black hole rotation, i.e., the sign of the angular momentum accretion rate is opposite to the sign of the black hole spin parameter.

In many respects our two-dimensional model of a thin disk of the Vlasov gas differs from spherically symmetric, three dimensional models known from \cite{Olivier1,Olivier2,cieslik}. We find, for instance, that the rest-mass accretion rate, normalized by the black hole mass and the asymptotic rest-mass surface density, increases with the asymptotic temperature. This stays in contrast to the spherically symmetric case, known for the Schwarzschild or Reissner-Nordstr\"{o}m metric, in which the rest-mass accretion rate, normalized by the square of the black-hole mass and the asymptotic rest-mass density, decreases with the asymptotic temperature. We should state that this behavior depends on the parameterization of solutions and is related to the properties of the two dimensional Maxwell-J\"{u}ttner distribution, assumed at infinity.

Equations governing the Vlasov gas at the equatorial plane of the Kerr spacetime can be derived in two frameworks. We will work in a 2+1 dimensional setup, based on the 3 dimensional metric induced at the equatorial plane. Alternatively, one can derive the same equations, working in the full 3+1 dimensional framework and imposing conditions which effectively restrict the motion to the equatorial plane. For clarity and completeness, this alternative formulation will be summarized in the appendix \ref{appendixA}.

The order of this paper is as follows. Section \ref{sec:preliminaries} contains mathematical preliminaries and conventions concerning the Kerr metric and the geodesic motion in the Kerr spacetime. In section \ref{sec:vlasov} we introduce our model of the disk of the collisionless Vlasov gas confined to the equatorial plane of the Kerr metric. We specify the distribution function and derive the expressions for the particle current surface density. Section \ref{sec:orbits} discusses the phase space of unbound orbits. In Section \ref{sec:current} we compute the particle current surface density and the rest-mass surface density of the disk. Section \ref{sec:accretionrates} is devoted to the rest-mass, energy, and angular momentum accretion rates; in particular, we derive analytic limits of these accretion rates for zero and infinite asymptotic temperatures of the gas. Section \ref{sec:discussion} contains a short discussion of the results.

\section{Preliminaries}
\label{sec:preliminaries}

\subsection{Metric conventions}

We use standard geometric units with $c = G = 1$, where $c$ denotes the speed of light, and $G$ is the gravitational constant. The signature of the metric is $(-,+,+,+)$.

We will work in Boyer-Lindquist coordinates $(t,r,\theta,\varphi)$. They are divergent at the black hole horizon, but, in comparison to more refined coordinate choices, they result in relatively simple formulas. In Boyer-Lindquist coordinates the Kerr metric can be written as
\begin{equation}
\label{kerr4d}
    g = g_{t t} dt^2 + 2 g_{t \varphi} dt d\varphi + g_{r r} dr^2 + g_{\theta \theta} d\theta^2 + g_{\varphi \varphi} d\varphi^2,
\end{equation}
where
\begin{subequations}
\begin{eqnarray}
g_{t t} & = & -1 + \frac{2Mr}{\rho^2}, \\
g_{t \varphi} & = & - \frac{2 M a r \sin^2 \theta}{\rho^2}, \\
g_{r r} & = & \frac{\rho^2}{\Delta}, \\
g_{\theta \theta} & = & \rho^2, \\
g_{\varphi \varphi} & = & \left( r^2 + a^2 + \frac{2 M a^2 r \sin^2 \theta}{\rho^2} \right) \sin^2 \theta,
\end{eqnarray}
\end{subequations}
and we denote
\begin{subequations}
\begin{eqnarray}
\Delta & = & r^2 - 2 M r + a^2, \\
\rho^2 & = & r^2 + a^2 \cos^2 \theta.
\end{eqnarray}
\end{subequations}
Note that many authors use the symbol $\Sigma$ in place of $\rho^2$ (cf.\ \cite{isco}). The contravariant metric component read
\begin{subequations}
\begin{eqnarray}
g^{t t} & = & - \frac{(r^2 + a^2)^2 - \Delta a^2 \sin^2 \theta}{\Delta \rho^2} \nonumber \\
& = & - \frac{1}{\Delta} \left( r^2 + a^2 + \frac{2 M r a^2 \sin^2 \theta}{\rho^2} \right), \\
g^{t \varphi} & = & - \frac{2 M r a}{\Delta \rho^2}, \\
g^{r r} & = & \frac{\Delta}{\rho^2}, \\
g^{\theta \theta} & = & \frac{1}{\rho^2}, \\
g^{\varphi \varphi} & = & \frac{1}{\Delta \rho^2} \left( -a^2 + \frac{\Delta}{\sin^2 \theta} \right).
\end{eqnarray}
\end{subequations}
The Kerr spacetime is characterized by the mass $M$ and the angular momentum $J = M a$. In general, we will restrict ourselves to the region outside the black hole horizon, with $\Delta > 0$.

The metric induced at the equatorial plane $\theta=\pi/2$ has the form
\begin{equation}
\label{kerr3d}
    \gamma = \gamma_{tt} dt^2 + 2 \gamma_{t \varphi} dt  d\varphi + \gamma_{rr}dr^2 + \gamma_{\varphi \varphi} d \varphi^2,
\end{equation}
where
\begin{subequations}
\begin{eqnarray}
    \gamma_{tt} & = & - 1 + \frac{2M}{r}, \\
    \gamma_{t \varphi} & = & - \frac{2 M a}{r}, \\
    \gamma_{rr} & = & \frac{r^2}{\Delta}, \\
    \gamma_{\varphi \varphi} & = & r^2 + a^2 \left( 1 + \frac{2M}{r} \right).
\end{eqnarray}
\end{subequations}
Contravariant components of $\gamma$ read
\begin{subequations}
\begin{eqnarray}
\gamma^{tt} & = & - \frac{1}{\Delta} \left( r^2 + a^2 + \frac{2 M a^2}{r} \right), \\
\gamma^{t \varphi} & = & - \frac{2 M a}{\Delta r}, \\
\gamma^{rr} & = & \frac{\Delta}{r^2}, \\
\gamma^{\varphi \varphi} & = & \frac{1}{\Delta} \left( 1 - \frac{2 M}{r} \right).
\end{eqnarray}
\end{subequations}
An easy calculation yields $\mathrm{det} \, g^{\mu \nu} = -1/(\rho^4 \sin^2{\theta}  )$ and $\mathrm{det} \, \gamma^{\mu\nu} = - 1/r^2$. Note that $\gamma$ is not
a solution of the vacuum Einstein equations in 2+1 dimensions. Such an equivalent of the Kerr solution does not exist, except for the so-called BTZ solution \cite{BTZ}, which assumes a negative cosmological constant.

\subsection{Time-like geodesics}

Time-like geodesic equations can be expressed in the Hamiltonian form
\begin{subequations}
\begin{eqnarray}
\frac{d x^\mu}{d \tau} & = & \frac{\partial H}{\partial p_\mu}, \\
\frac{d p_\nu}{d \tau} & = & - \frac{\partial H}{\partial x^\nu},
\end{eqnarray}
\end{subequations}
where $p^\mu = d x^\mu / d \tau$, $H(x^\alpha,p_\beta) = \frac{1}{2} g^{\mu\nu}(x^\alpha) p_\mu p_\nu = - \frac{1}{2} m^2$, and $m$ denotes the particle rest mass. By standard arguments, $H$, $p_t \equiv - E$ and $p_\varphi \equiv l_z$ are constants of motion. The fourth constant---the so-called Carter constant $l$---follows from the separation of variables in the Hamilton-Jacobi equation
\begin{equation}
    \frac{\partial S}{\partial \tau} + H(x^\alpha, \partial_\beta S) = 0.
\end{equation}
The ansatz
\begin{equation}
    S = - H \tau + W = \frac{1}{2} m^2 \tau + W 
\end{equation}
with
\begin{eqnarray*}
    W & = & \int p_t dt + \int p_\varphi d \varphi + \int p_r dr + \int p_\theta d \theta  \nonumber \\
    & = & - E t + l_z \varphi + W_r(r) + W_\theta(\theta)
\end{eqnarray*}
yields the ``time-independent'' Hamilton--Jacobi equation
\begin{equation}
    g^{\mu \nu} \partial_\mu W \partial_\nu W = - m^2
\end{equation}
or, in explicit terms,
\begin{widetext}
\begin{equation}
    - \frac{(r^2 + a^2)^2 - \Delta a^2 \sin^2 \theta}{\Delta} E^2 + \frac{4 M r a}{\Delta} E l_z 
    + \Delta (W_r^\prime)^2 + (W_\theta^\prime)^2 + \frac{1}{\Delta} \left( -a^2 + \frac{\Delta}{\sin^2 \theta} \right) l_z^2  = - m^2 \rho^2.
\end{equation}
Rearranging terms, we get
\begin{equation}
\label{carter_separation}
    (W_\theta^\prime)^2 + m^2 a^2 \cos^2 \theta + \left( \frac{l_z}{\sin \theta} - a \sin \theta E \right)^2 = - m^2 r^2 - \Delta (W_r^\prime)^2 + \frac{1}{\Delta} \left[ (r^2 + a^2) E - a l_z \right]^2.
\end{equation}
\end{widetext}
Since the left-hand side depends only on $\theta$, while the right-hand side depends only on $r$, both sides have to be constant. We denote this constant value by $l^2$ (note that the left-hand side of Eq.\ (\ref{carter_separation}) is explicitly non-negative). Thus
\begin{equation}
\label{W_theta}
    (W_\theta^\prime)^2 \equiv \Theta(\theta) = l^2 - m^2 a^2 \cos^2 \theta - \left( \frac{l_z}{\sin \theta} - a \sin \theta E \right)^2,
\end{equation}
and
\begin{equation}
\label{W_r}
    \Delta^2 (W_r^\prime)^2 \equiv R(r) = \left[ (r^2 + a^2) E - a l_z \right]^2 - \Delta (m^2 r^2 + l^2).
\end{equation}
The equations of motion can be integrated following the standard procedure, i.e., by computing the derivatives $\partial S/\partial m$, $\partial S/\partial E$, $\partial S/\partial l_z$, $\partial S/\partial l$, and equating the results to a new set of constants $\beta_0$, $\beta_1$, $\beta_2$, $\beta_3$ (say) \cite{chandrasekhar,landaulifszyc}.


\subsection{Time-like geodesics at the equatorial plane}

The description of time-like geodesics restricted to the equatorial plane ($\theta = \pi/2$) is much simpler. We have $p_\theta = 0$, $W_\theta = 0$. Equation \eqref{W_theta} reduces to
\begin{equation}
    l^2 = (l_z - a E)^2.
\end{equation}
We will assume a convention with $l \ge 0$. Thus
\begin{equation}
    \epsilon_\sigma l = l_z - a E,
\end{equation}
where $\epsilon_\sigma = \pm 1$. Consequently
\begin{eqnarray}
    R(r) & = & (r^2 E - \epsilon_\sigma a l)^2 - \Delta (m^2 r^2 + l^2) \nonumber \\
    & = & r^4 \left[ E^2 - \left( 1 - \frac{2M}{r} \right) \left( m^2 + \frac{l^2}{r^2} \right) \right. \nonumber \\
    && \left. - \frac{2 \epsilon_\sigma a E l + a^2 m^2}{r^2} \right].
\end{eqnarray}
Momentum components associated with the geodesic motion confined to the equatorial plane can be written as
\begin{subequations}
\begin{eqnarray}
p_t & = & - E, \\
p_r & = & \epsilon_r \frac{\sqrt{R}}{\Delta}, \label{pr} \\
p_\theta & = & 0, \\
p_\varphi & = & l_z = \epsilon_\sigma l + a E = \epsilon_\sigma (l + \epsilon_\sigma a E),
\end{eqnarray}
\end{subequations}
where $\epsilon_r = \pm 1$ corresponds to the radial direction of motion.

Alternatively, one can obtain equations governing the motion confined to the equatorial plane directly from the Hamiltonian
\begin{equation}
\label{hequat}
    H = \frac{1}{2} \left( \gamma^{tt} p_t^2 + 2 \gamma^{t\varphi} p_t p_\varphi + \gamma^{rr} p_r^2 + \gamma^{\varphi \varphi} p_\varphi^2 \right).
\end{equation}
Here again, $p_t = -E$, $p_\varphi = l_z$, and $H = - \frac{1}{2}m^2$ have to be constant. Solving the equation $H = -\frac{1}{2}m^2$ for $p_r$ yields Eq.\ (\ref{pr}).

Following \cite{Olivier1,Olivier3}, we use dimensionless variables defined as
\begin{equation}
\label{dimensionless}
    \alpha = \frac{a}{M}, \quad \varepsilon = \frac{E}{m}, \quad \lambda = \frac{l}{M m}, \quad \xi = \frac{r}{M}.
\end{equation}
Thus,
\begin{equation}
    R(r) = M^4 m^2 \tilde R(\xi),
\end{equation}
where
\begin{eqnarray}
\label{Rtilde}
    \tilde R(\xi) & = & (\xi^2 \varepsilon - \epsilon_\sigma \alpha \lambda)^2 - (\xi^2 - 2 \xi + \alpha^2)(\xi^2 + \lambda^2) \\
    & = & \xi^4 \left[ \varepsilon^2 - \left( 1 - \frac{2}{\xi} \right) \left( 1 + \frac{\lambda^2}{\xi^2} \right) - \frac{2 \epsilon_\sigma \alpha \varepsilon \lambda + \alpha^2}{\xi^2}\right], \nonumber
\end{eqnarray}
and
\begin{subequations}
\begin{eqnarray}
p_t & = & - E = - m \varepsilon, \\
p_r & = & \epsilon_r \frac{\sqrt{R}}{\Delta} = \frac{\epsilon_r m \sqrt{\tilde R}}{\xi^2 - 2 \xi + \alpha^2}, \\
p_\theta & = & 0, \\
p_\varphi & = & l_z = \epsilon_\sigma l + a E = M m (\epsilon_\sigma \lambda + \alpha \varepsilon) \nonumber \\
& = & M m \epsilon_\sigma (\lambda + \epsilon_\sigma \alpha \varepsilon).
\end{eqnarray}
\end{subequations}

In this work we will generally parametrize geodesics using the Carter constant $l$ (or $\lambda$) instead of $l_z$. This is, of course, a matter of preference. Our choice was partially motivated by a similar substitution in Eq.\ (3) of \cite{Olivier3}; it also simplifies a few algebraic relations used in this paper. We should emphasize that the sign $\epsilon_\sigma$ is associated with the quantity $l_z - aE$ and not with $l_z$. In particular $l_z = 0$ implies $l = - \epsilon_\sigma a E$. On the other hand in many important cases the signs of $l_z$ and $\epsilon_\sigma$ coincide. Thus, we will generally refer to trajectories with $\epsilon_\sigma \alpha > 0$ as prograde ones, and to those satisfying $\epsilon_\sigma \alpha < 0$ as retrograde ones.

There are three well-known types of circular orbits at the equatorial plane, two of which will play an important role in the subsequent analysis \cite{isco,dewitt}. There is the circular photon orbit with the radius
\begin{equation}
\label{xiphoton}
    \xi_\mathrm{ph} = 2 + 2 \cos \left[ \frac{2}{3} \mathrm{arccos} \left( - \epsilon_\sigma  \alpha \right) \right],
\end{equation}
a solution of $(\xi-3)\sqrt{\xi} + 2 \epsilon_\sigma \alpha = 0$ or $(\xi - 3)^2 \xi - 4 \alpha^2 = 0$. The circular photon orbit corresponds to a null geodesic, but it appears naturally as a limit of time-like orbits. There is the so-called marginally bound orbit with the dimensionless radius
\begin{equation}
\label{ximb}
    \xi_\mathrm{mb} = 2 - \epsilon_\sigma \alpha + 2 \sqrt{1-\epsilon_\sigma \alpha},
\end{equation}
which is a solution to $\xi(\xi-4) + 4 \alpha \epsilon_\sigma \sqrt{\xi} - \alpha^2=0$ (see also Eq.\ \ref{xi4_lambda}). For unbound orbits $\xi_\mathrm{mb}$ is related to the so-called fly-by radius \cite{Szybka4}. The dimensionless radius of the innermost stable circular orbit (marginally stable orbit, ISCO) is given by
\begin{equation}
\label{isco_formula}
\xi_\text{ms} = 3 + Z_2 - \epsilon_\sigma \alpha \sqrt{\frac{(3 - Z_1)(3 + Z_1 + 2 Z_2)}{\alpha^2}},
\end{equation}
where $Z_1 = 1+ \sqrt[3]{1-\alpha^2}(\sqrt[3]{1+\epsilon_\sigma\alpha} + \sqrt[3]{1-\epsilon_\sigma\alpha})$, $Z_2 = \sqrt{3 \alpha^2+Z_1^2}$ \cite{isco}. The radius $\xi_\mathrm{ms}$ is a solution of the equation $(\xi -6) \xi + 8 \alpha \sqrt{\xi} - 3 \alpha ^2 = 0$. The location of the innermost stable circular orbit does not seem to play a crucial role in our analysis. However, the related polynomial in $\sqrt{\xi}$, as well as those defining $\xi_\text{ph}$ and $\xi_\text{mb}$, do appear in the integrands of our formulas for accretion rates. 

\section{Vlasov equation at the equatorial plane}
\label{sec:vlasov}

The Vlasov gas considered in this paper is a collection of free particles moving along time-like future-directed geodesic lines. It is described in terms of the distribution function $f = f(x^\mu,p_\nu)$, defined on the one-particle phase space. For collisionless particles, the Vlasov equation can be expressed as a requirement that the probability function $f$ is constant along a geodesic:
\begin{eqnarray}
\label{vlasoveq}
    \frac{df}{d\tau} & = & \frac{\partial f}{\partial x^\mu} \frac{d x^\mu}{d \tau} + \frac{\partial f}{\partial p_\nu} \frac{d p_\nu}{d \tau} = \frac{\partial f}{\partial x^\mu} \frac{\partial H}{\partial p_\mu} - \frac{\partial f}{\partial p_\nu} \frac{\partial H}{\partial x^\nu} \nonumber \\
    & = & \{ H, f \} = 0.
\end{eqnarray}
Here $\{ \cdot, \cdot \}$ denotes the Poisson bracket.

A formal solution of the above equation has been obtained in the context of the Kerr metric in \cite{zannias}. We will repeat this reasoning here, assuming that the motion is restricted to the equatorial plane. In this case, $(x^\mu,p_\nu) = (t,r,\varphi,p_t,p_r,p_\varphi)$, and the Hamiltonian $H$ is given by Eq.\ (\ref{hequat}). The idea is based on a suitable canonical (symplectic) transformation of the phase-space variables $(t,r,\varphi,p_t,p_r,p_\varphi) \mapsto (Q^\mu,P_\nu)$. It is given in terms of a generating function (the so-called abbreviated action)
\begin{eqnarray}
    W & = & W(t,r,\varphi,m,E,l_z) = \int p_t dt + \int p_r dr + \int p_\varphi d \varphi \nonumber \\
    & = & -E t + l_z \varphi + W_r.
\end{eqnarray}
We choose new momenta in the form
\begin{equation}
    P_0 = m, \quad P_1 = E, \quad P_2 = l_z.
\end{equation}
The corresponding conjugate variables $Q^\mu$ are defined as $Q^\mu = \partial W/\partial P_\mu$, i.e., 
\begin{equation}
    Q^0 = \frac{\partial W}{\partial m}, \quad Q^1 = \frac{\partial W}{\partial E}, \quad Q^2 = \frac{\partial W}{\partial l_z}.
\end{equation}
The Vlasov equation (\ref{vlasoveq}), written in terms of the Poisson bracket, is covariant with respect to canonical transformations. Since $H = - \frac{1}{2}P_0^2$, it reads, in the new variables $(Q^\mu,P_\nu)$,
\begin{equation}
\label{vlasovglobal}
    - P_0 \frac{\partial f}{\partial Q^0} = 0.
\end{equation}
Consequently, any distribution function of the form
\begin{equation}
\label{general solution}
    f = f(Q^1,Q^2,P_0,P_1,P_2)
\end{equation}
satisfies the Vlasov equation. Further restrictions on the form of the distribution function can be obtained by imposing symmetry conditions.

The distribution function $f$ is not an observable quantity on its own. We define the components of the of the particle current surface density as
\begin{equation}
\label{jmu3d}
    J_\mu(x) = \int f(x,p) p_\mu \sqrt{- \mathrm{det} \, \gamma^{\mu\nu}} d p_t d p_r d p_\varphi,
\end{equation}
where $\mu = t, r, \varphi$, and $f = f(t,r,\varphi,p_t,p_r,p_\varphi)$. Similarly, the surface energy-momentum tensor components are
\begin{equation}
    T_{\mu \nu} (x) = \int f(x,p) p_\mu p_\nu \sqrt{- \mathrm{det} \, \gamma^{\mu\nu}} d p_t d p_r d p_\varphi.
\end{equation}
A covariant particle number surface density $n_s$ can be defined as
\begin{equation}
\label{n_s}
    n_s = \sqrt{- \gamma_{\mu \nu} J^\mu J^\nu}.
\end{equation}

We consider a model of a stationary accretion disk consisting of same-mass particles, confined to the equatorial plane and extending to infinity. At infinity the gas is assumed to be essentially ``at rest'' and to be characterized by a constant particle number surface density $n_{s,\infty}$. More precisely, we require that at infinity the distribution function should be given by the two dimensional Maxwell-J\"{u}ttner distribution \cite{juttner1,juttner2}. On the other hand, the motion of the gas near the black hole should be governed by its gravitational field. A distribution function fulfilling these assumptions can be chosen as
\begin{eqnarray}
    f & = & A \delta \left( \sqrt{-p_\mu p^\mu} - m_0 \right) \exp \left( \frac{\beta}{m_0} p_t \right) \nonumber \\
    & = & A \delta(P_0 - m_0) \exp\left( - \frac{\beta}{m_0} P_1 \right),
\label{f3d}    
\end{eqnarray}
where $m_0$ (the particle mass) and $A$ are constants. Here $\beta = m_0/(k_\mathrm{B} T)$, $T$ is the asymptotic temperature of the gas, and $k_\mathrm{B}$ denotes the Boltzmann constant. The factor $\delta \left( \sqrt{-p_\mu p^\mu} - m_0 \right)$ ensures that the momenta satisfy the mass shell condition. Here, and in what follows, we only take into account future-pointing momenta. Clearly, the distribution function given by Eq. (\ref{f3d}) is of the form \eqref{general solution}; it satisfies the collisionless Vlasov equation (\ref{vlasovglobal}) and coincides with the standard Maxwell-J\"{u}ttner distribution at the infinity.

In what follows, we will also use the rest-mass surface density, defined as
\begin{equation}
    \rho_s = m_0 n_s.
\end{equation}

The components of the particle current surface density $J_t$, $J_r$, and $J_\varphi$ can be computed according to Eqs.\ (\ref{jmu3d}). To evaluate the integrals over momenta appearing in Eqs.\ (\ref{jmu3d}), one needs to control the domain in the phase space occupied by relevant particle trajectories---this is probably the most difficult part of the calculation. The coordinate transformations introduced in the remainder of this section and the analysis of Sec.\ \ref{sec:orbits} are devoted to this problem.

We will only take into account unbound trajectories, i.e., trajectories that reach infinity. Bound trajectories could also play a role, but since collisions between gas particles are neglected and we are working on a fixed spacetime, contributions corresponding to bound and unbound trajectories can be treated separately. Thus, for simplicity, we assume that no bound trajectories are present. The reader interested in the phase-space mixing occurring for bound trajectories in the equatorial plane of the Kerr spacetime is referred to \cite{Olivier3}.

Unbound trajectories are naturally divided into two classes. There are orbits characterized by relatively small values of angular momentum, which end in the black hole. We refer to such orbits as absorbed ones. The second class consists of orbits with sufficiently high angular momentum, which are scattered off the centrifugal barrier. They originate and end at infinity. We will refer to those orbits as scattered trajectories. Even though we are mainly interested in the accretion process, both types of unbound trajectories have to be taken into account, as they all contribute to the total ``budget'' of trajectories at infinity, where the Maxwell-J\"{u}ttner distribution is assumed.

The analysis of unbound trajectories is simpler in terms of conserved variables such as $m$, $E$, $l_z$ (or $l$), instead of momenta $p_t$, $p_r$, $p_\varphi$. Thus, we change momentum variables $p_t$, $p_r$, $p_\varphi$ to
\begin{subequations}
\begin{eqnarray}
m^2 & = & - \left( \gamma^{tt} p_t^2 + 2 \gamma^{t \varphi} p_t p_\varphi + \gamma^{rr} p_r^2 + \gamma^{\varphi \varphi} p_\varphi^2 \right), \\
E & = & - p_t, \\
l & = & \epsilon_\sigma(p_\varphi + a p_t).
\end{eqnarray}
\end{subequations}
The Jacobian of the above transformation reads
\begin{equation}
    \frac{\partial (m^2, E, l)}{\partial (p_t,p_r,p_\varphi)} = - \frac{2 \epsilon_\sigma p_r \Delta}{r^2}.
\end{equation}
Consequently,
\begin{equation}
    dp_t dp_r dp_\varphi = \frac{r^2}{2 |p_r| \Delta} d m^2 dE d l = \frac{r^2 m}{|p_r| \Delta} dm dE dl,
\end{equation}
where $|p_r| \Delta = \sqrt{R(r)}$.

In terms of dimensionless quantities (\ref{dimensionless}) the momentum integration element can be written as
\begin{equation}
    \sqrt{- \mathrm{det} \, \gamma^{\mu \nu}} dp_t dp_r dp_\varphi = \frac{\xi m^2 dm d \varepsilon d \lambda}{\sqrt{\tilde R}}.
\end{equation}
This yields formal expressions for the components $J_\mu$ in the form
\begin{widetext}
\begin{subequations}
\label{Jmugeneral3d}
\begin{eqnarray}
J_t & = & - A m_0^3 \xi \sum_{\epsilon_\sigma = \pm 1} \int \exp(- \beta \varepsilon) \varepsilon \frac{d \varepsilon d\lambda}{\sqrt{\tilde R}}, \\
J_r & = & \frac{A M^2 m_0^3 \xi}{\Delta} \sum_{\epsilon_\sigma = \pm 1} \int \epsilon_r \exp(- \beta \varepsilon) d \varepsilon d \lambda, \\
J_\varphi & = & A M m_0^3 \xi \sum_{\epsilon_\sigma = \pm 1} \int \exp(- \beta \varepsilon) ( \epsilon_\sigma \lambda + \alpha \varepsilon) \frac{d \varepsilon d \lambda}{\sqrt{\tilde R}}.
\end{eqnarray}
\end{subequations}
In the following we will also need explicit expressions for
\begin{equation}
\label{jrupgeneral}
    J^r = \frac{A m_0^3}{\xi} \sum_{\epsilon_\sigma = \pm 1} \int \epsilon_r \exp( -\beta \varepsilon ) d \varepsilon d \lambda
\end{equation}
and
\begin{eqnarray}
    T\indices{^r_t} & = & - \frac{A m_0^4}{\xi} \sum_{\epsilon_\sigma = \pm 1} \int \epsilon_r \exp( - \beta \varepsilon ) \varepsilon d \varepsilon d \lambda, \\
     T\indices{^r_\varphi} & = & \frac{A m_0^4 M}{\xi} \sum_{\epsilon_\sigma = \pm 1} \int \epsilon_r \exp( - \beta \varepsilon ) (\epsilon_\sigma \lambda + \alpha \varepsilon ) d \varepsilon d \lambda.
\end{eqnarray}
\end{widetext}

There is a simple test of correctness that can be performed at this point---one can compute the component $J_t$ in the flat Minkowski space limit $M \to 0$, $a \to 0$. This has to be done with some caution. In particular the mass $M$ in Eqs.\ (\ref{dimensionless}) defining the dimensionless quantities should be replaced with any positive constant with the mass dimension. The expression for $\tilde R$ yields, in the Minkowski limit,
\begin{equation}
   \nonumber
    \tilde R  \to \xi^4 \left( \varepsilon^2 -  1  - \frac{\lambda^2}{\xi^2}  \right),
\end{equation}
and hence
\begin{equation}
   \nonumber
    J_t = - \frac{A m_0^3}{\xi} \sum_{\epsilon_\sigma = \pm 1} \int \exp(- \beta \varepsilon) \varepsilon \frac{d \varepsilon d\lambda}{\sqrt{ \varepsilon^2 - 1 - \frac{\lambda^2}{\xi^2}}}.
\end{equation}
The key point is to control the range in the phase space over which the above integral is performed. The trajectories are straight lines with $\varepsilon > 1$ and $\lambda < \xi \sqrt{\varepsilon^2 - 1}$. In addition, a set of values $\varepsilon$, $m$, and $\lambda$, correspond to 2 possible values of the radial momentum $p_r = \pm |p_r|$. This gives $J_t$ of the form
\begin{equation}
    J_t = - \frac{4 A m_0^3}{\xi} \int_1^\infty d \varepsilon \exp (- \beta \varepsilon) \varepsilon \int_0^{\xi \sqrt{\varepsilon^2 - 1}} \frac{d \lambda}{\sqrt{\varepsilon^2 - 1 - \frac{\lambda^2}{\xi^2}}}.
\end{equation}
The last integral, over $\lambda$, is simply $\pi \xi/2$. Thus
\begin{eqnarray}
    J_t & = & - 2 \pi A m_0^3 \int_1^\infty d \varepsilon \exp (- \beta \varepsilon) \varepsilon  \nonumber \\
    & = & - 2 \pi A m_0^3 \frac{e^{-\beta}(1 + \beta)}{\beta^2}.
\label{jtlimit}
\end{eqnarray}
This result agrees with the standard computation of $J_t$ based on Cartesian coordinates. For completeness, we report this computation in Appendix \ref{sec:minkowski}. Notably formula (\ref{jtlimit}) agrees with Eq.\ (\ref{jtflat})
and formula (117) of \cite{Acuna}, which uses the Bessel function
\begin{equation}
K_{3/2}(\beta) = \sqrt{\frac{\pi}{2 \beta}} \frac{e^{-\beta}(1 + \beta)}{\beta}.
\end{equation}

The above result allows us also to write the expression for the asymptotic rest-mass surface density of matter at the equatorial plane
\begin{equation}
   \rho_{s,\infty} \equiv m_0 n_{s,\infty} = - m_0 J_t|_\infty = 2 \pi A m_0^4 \frac{1 + \beta}{\beta^2} \exp(- \beta).
\end{equation}
Hence, the factor $A$ in the integral expressions \eqref{Jmugeneral3d} for $J_\mu$ can be replaced by
\begin{equation}
\label{A}
    A = \frac{ \rho_{s,\infty}}{2 \pi m_0^4} \frac{\beta^2}{1 + \beta} \exp(\beta).
\end{equation}

\section{Classification of orbits}
\label{sec:orbits}

\begin{figure}[t]
\includegraphics[width=\columnwidth]{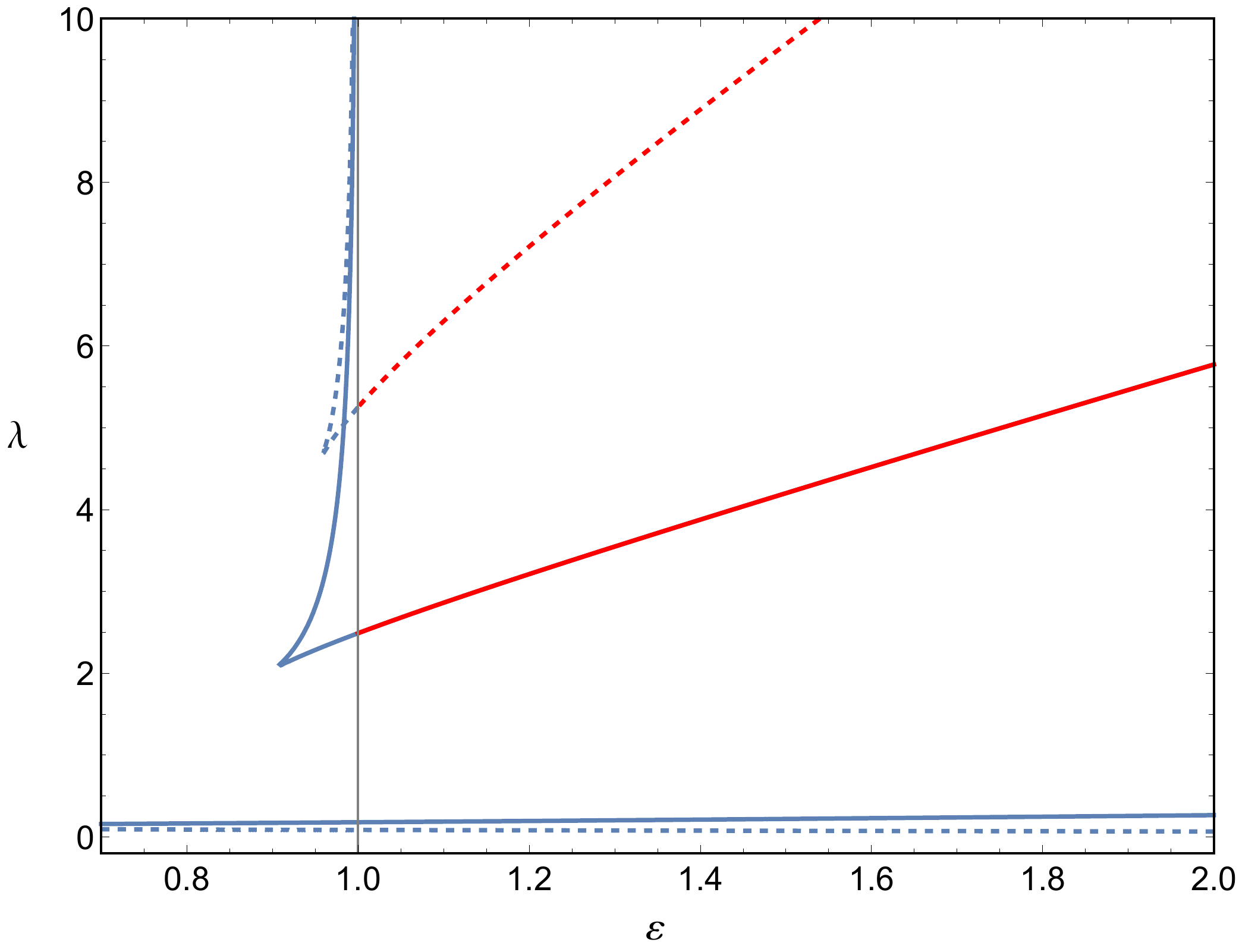}
\caption{\label{fig:groebner} Sample solutions of Eq.\ (\ref{groebner}) for $\alpha = 2/3$, $\epsilon_\sigma = +1$ (continuous line), $\epsilon_\sigma = -1$ (dashed line). Physically relevant branches, given by the parametric formula (\ref{parametric}), are depicted with red lines. Note the existence of mirrored branches in the remaining three quadrants of the full $\varepsilon-\lambda$ plane (not shown). }
\end{figure}

\begin{figure}[t]
\includegraphics[width=\columnwidth]{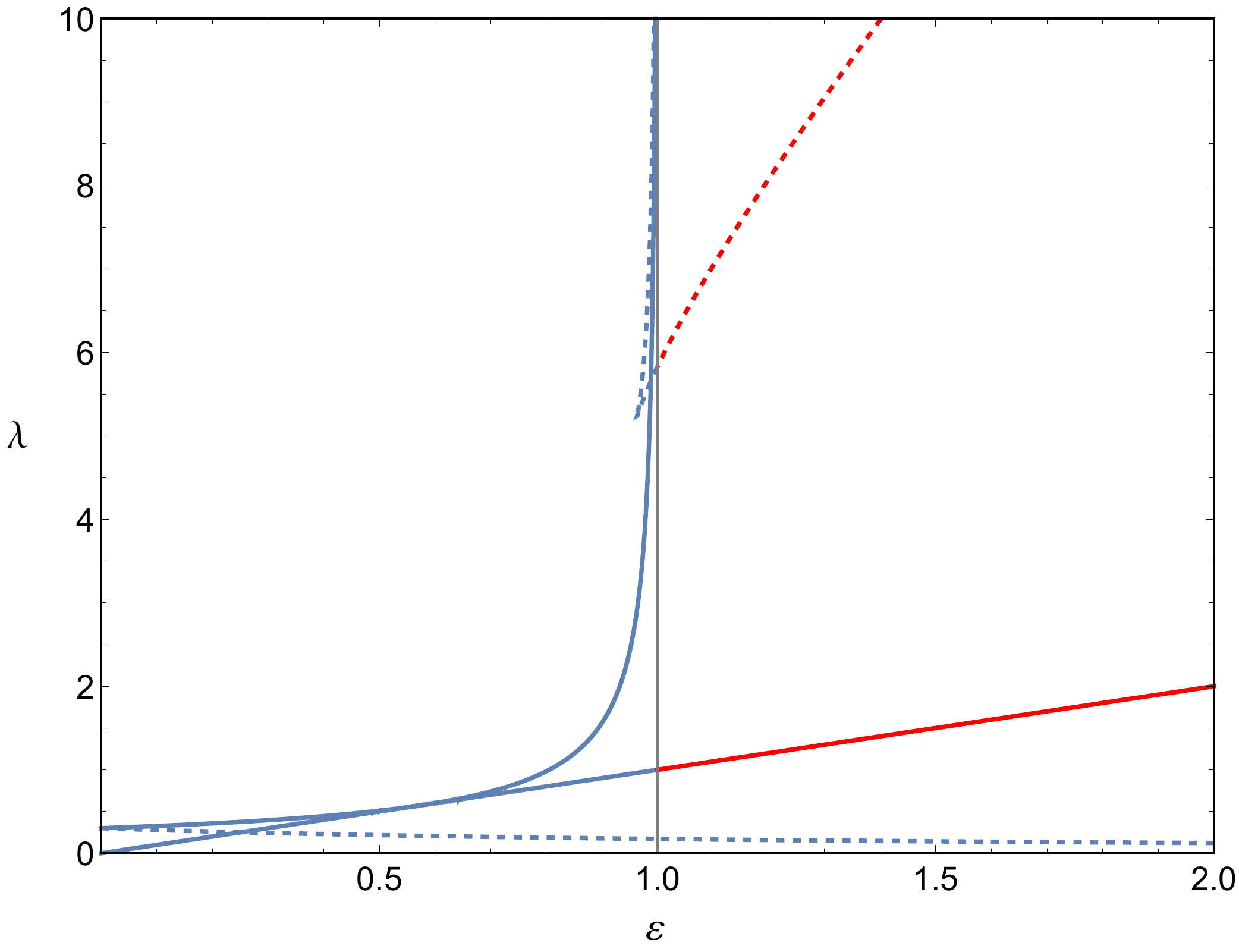}
\caption{\label{fig:groebner3} 
Sample solutions of Eq.\ (\ref{groebner}) for $\alpha = 1$, $\epsilon_\sigma = +1$ (continuous line), $\epsilon_\sigma = -1$ (dashed line). Physically relevant branches are depicted with red lines. For $\alpha = 1$ and $\epsilon_\sigma = +1$, $\lambda_c(\varepsilon,\epsilon_\sigma = +1) = \varepsilon$.}
\end{figure}

\begin{figure}[t]
\includegraphics[width=\columnwidth]{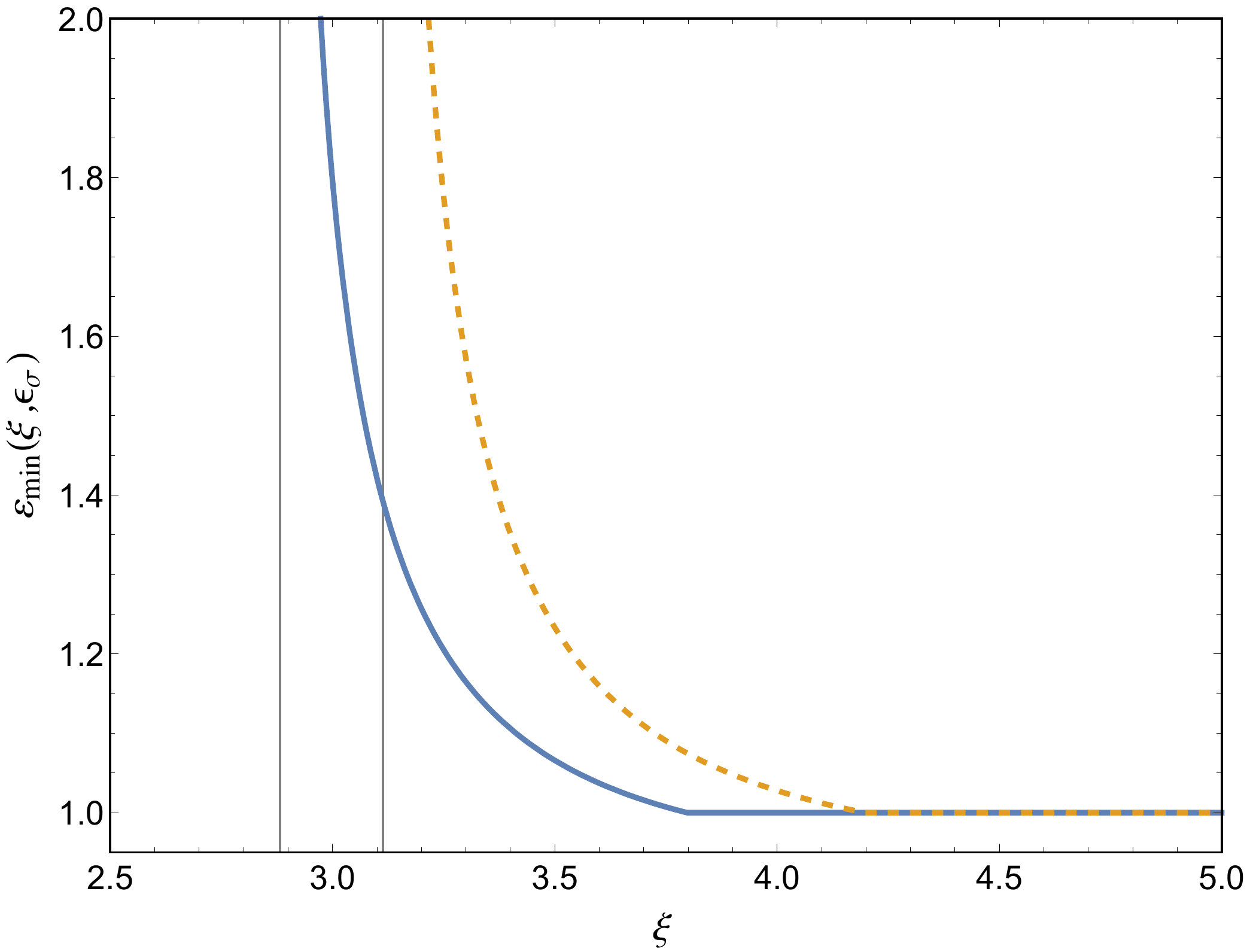}
\caption{\label{fig:emin} Sample graphs of $\varepsilon_\mathrm{min}(\xi,\epsilon_\sigma)$ for $\alpha = 1/10$ and $\epsilon_\sigma = +1$ (blue continuous line) and $\epsilon_\sigma = -1$ (orange dashed line). Vertical lines denote locations of circular photon orbits $\xi_\mathrm{ph}$ for $\epsilon_\sigma = \pm 1$. Note that $\varepsilon_\mathrm{min}(\xi,+1)$ is regular at $\xi_\mathrm{ph}$ corresponding to $\epsilon_\sigma = -1$.}
\end{figure}

\begin{figure}
\centering
\includegraphics[width=\columnwidth]{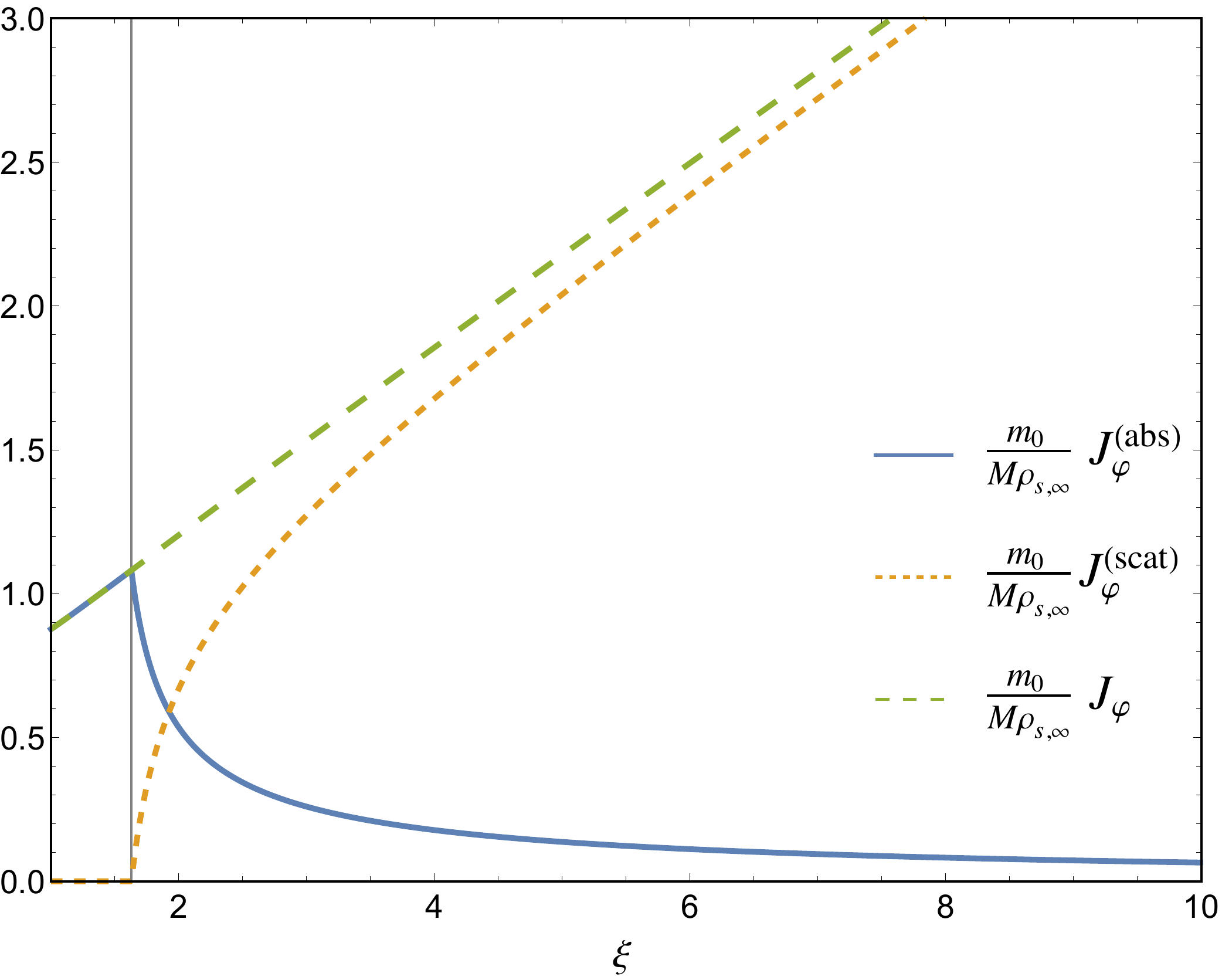}
\caption{\label{fig:jmu1} Components of the particle current surface density $J_{\varphi}$, $J^\mathrm{(abs)}_{\varphi}$, and $J^\mathrm{(scat)}_{\varphi}$ corresponding to $\epsilon_\sigma = +1$. The spin parameter $\alpha = 7/8$; $\beta = 0.1$. The vertical line marks the location of the circular photon orbit $\xi_\mathrm{ph}$.}
\end{figure}

\begin{figure}
\centering
\includegraphics[width=\columnwidth]{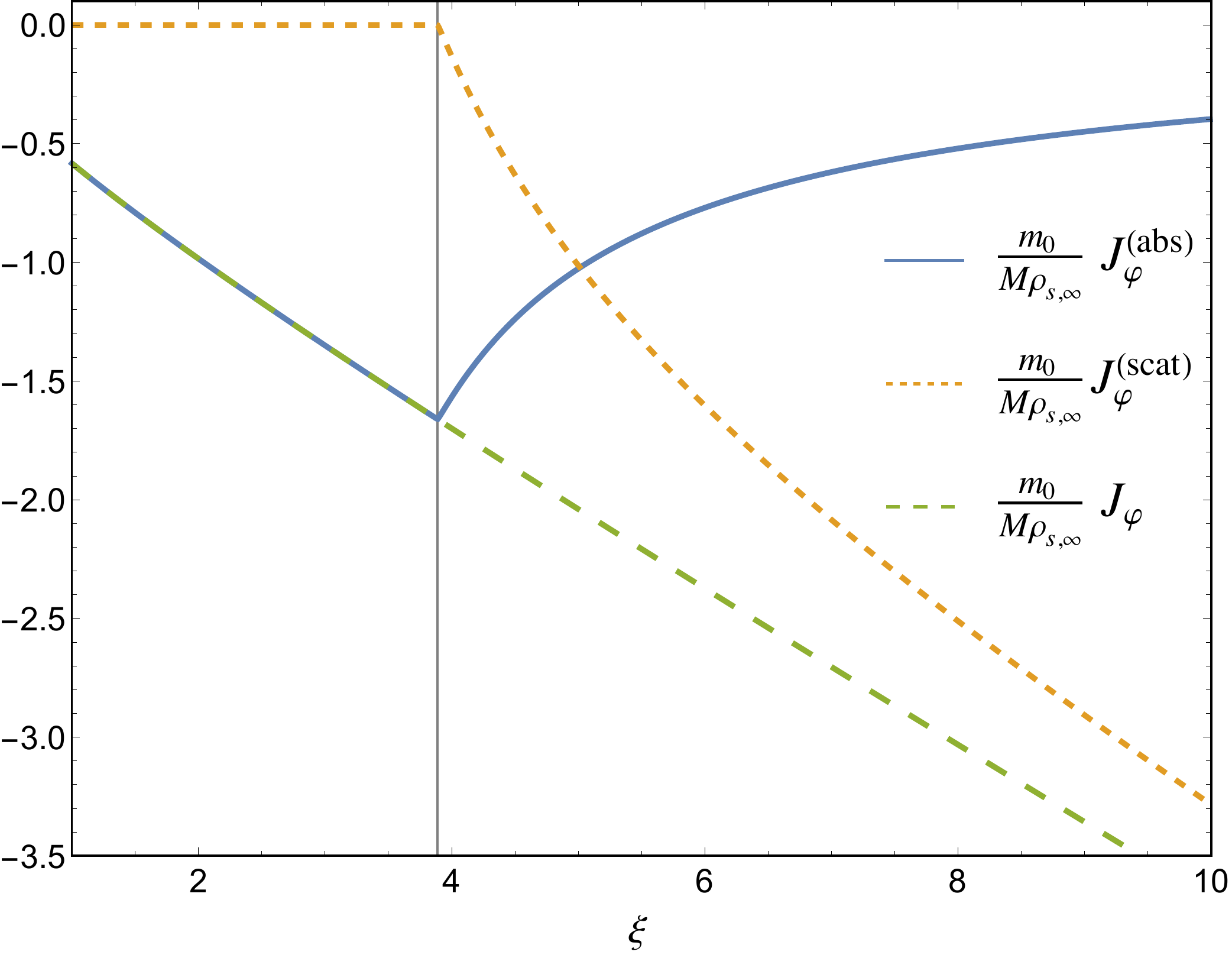}
\caption{\label{fig:jmu2} Same as in Fig.\ \ref{fig:jmu1}, but for $\epsilon_\sigma = -1$.}
\end{figure}

\begin{figure}
\centering
\includegraphics[width=\columnwidth]{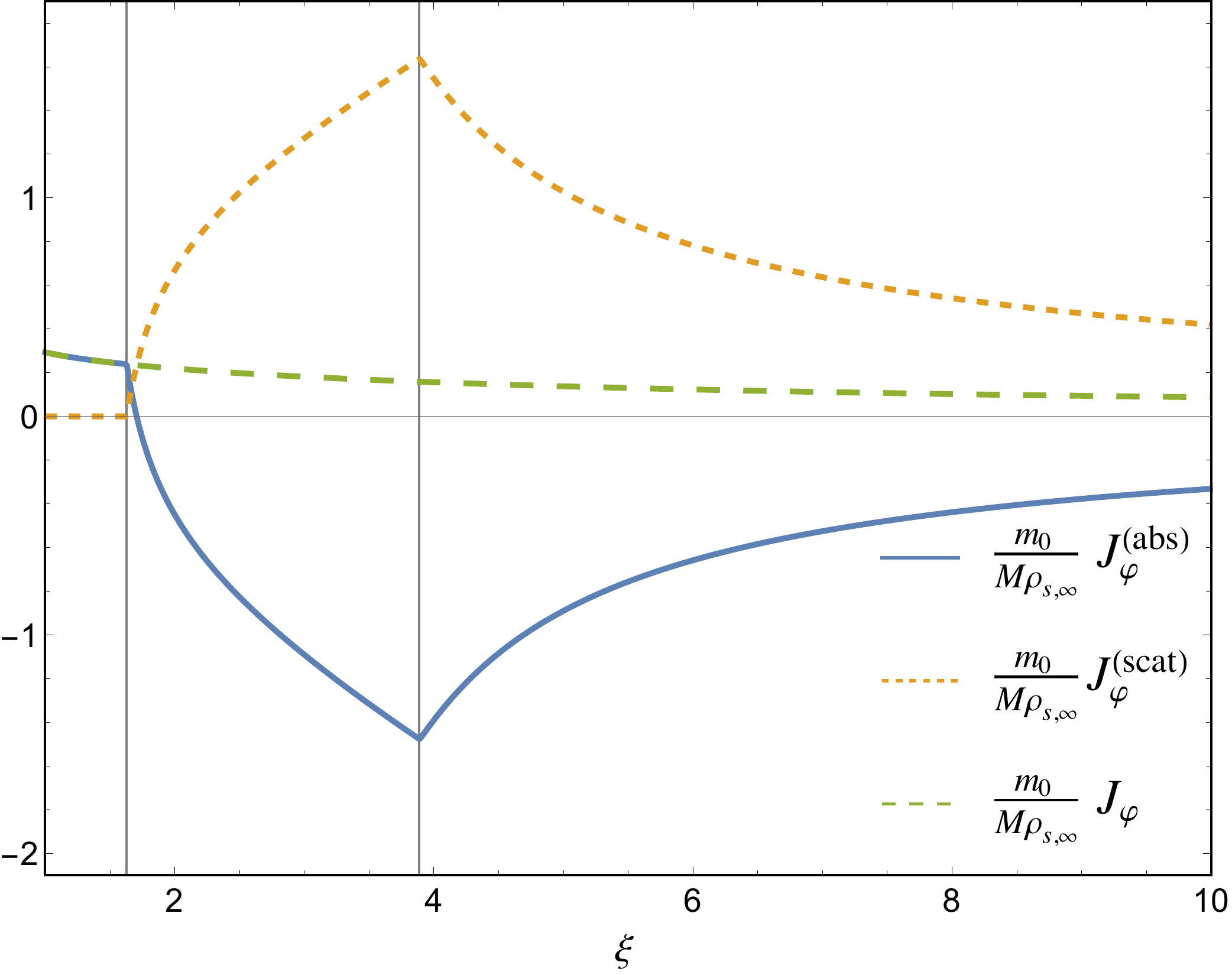}
\caption{\label{fig:jmu3}
Components of the particle current surface density $J_{\varphi}$, $J^\mathrm{(abs)}_{\varphi}$, and $J^\mathrm{(scat)}_{\varphi}$ obtained by summing the parts corresponding to $\epsilon_\sigma = \pm 1$. The spin parameter $\alpha = 7/8$; $\beta = 0.1$.}
\end{figure}

\begin{figure}
\centering
\includegraphics[width=\columnwidth]{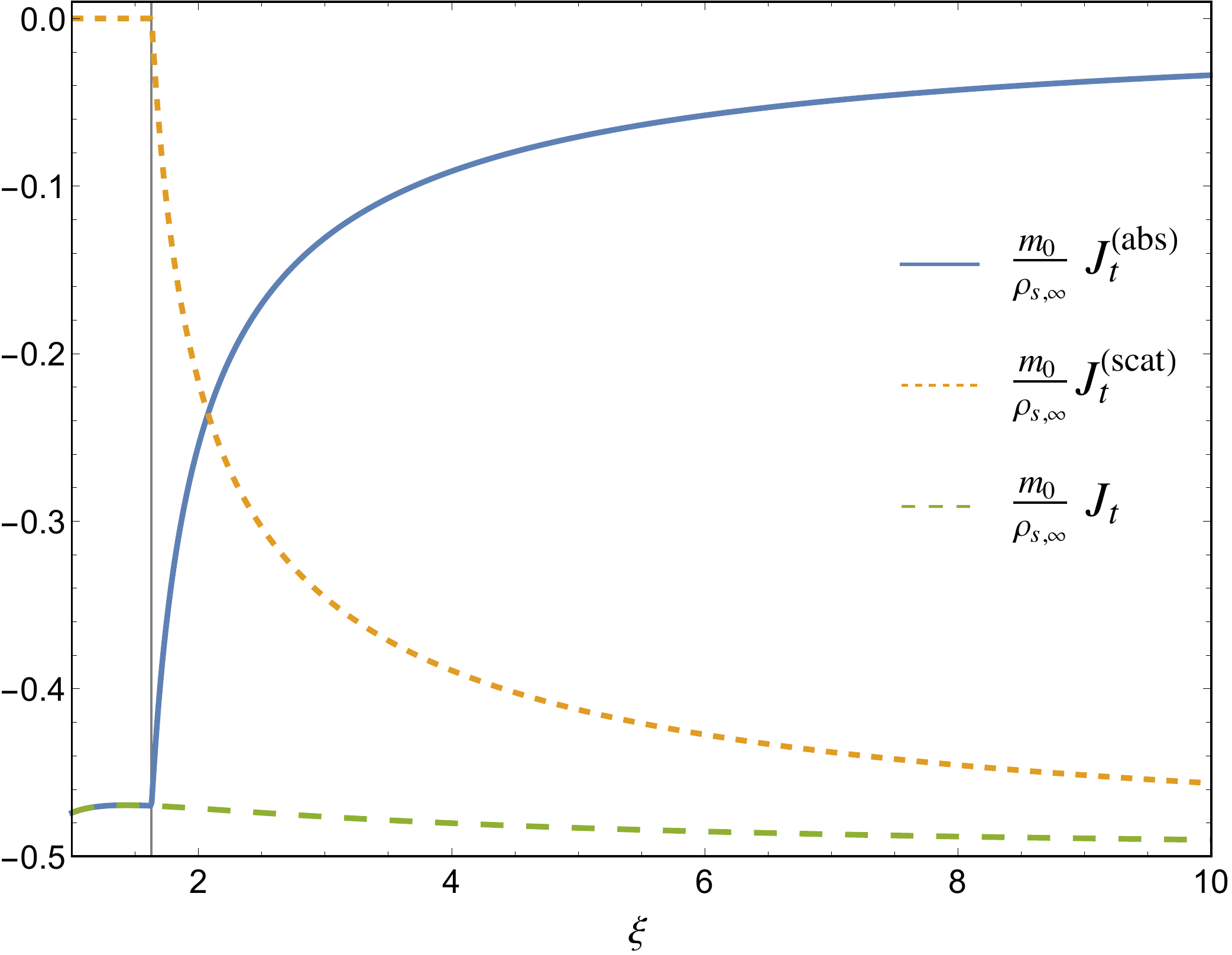}
\caption{\label{fig:jt1} Components of the particle current surface density $J_{t}$, $J^\mathrm{(abs)}_{t}$, and $J^\mathrm{(scat)}_{t}$ corresponding to $\epsilon_\sigma = +1$. The spin parameter $\alpha = 7/8$; $\beta = 0.1$. The vertical line marks the location of the circular photon orbit $\xi_\mathrm{ph}$.}
\end{figure}

\begin{figure}
\centering
\includegraphics[width=\columnwidth]{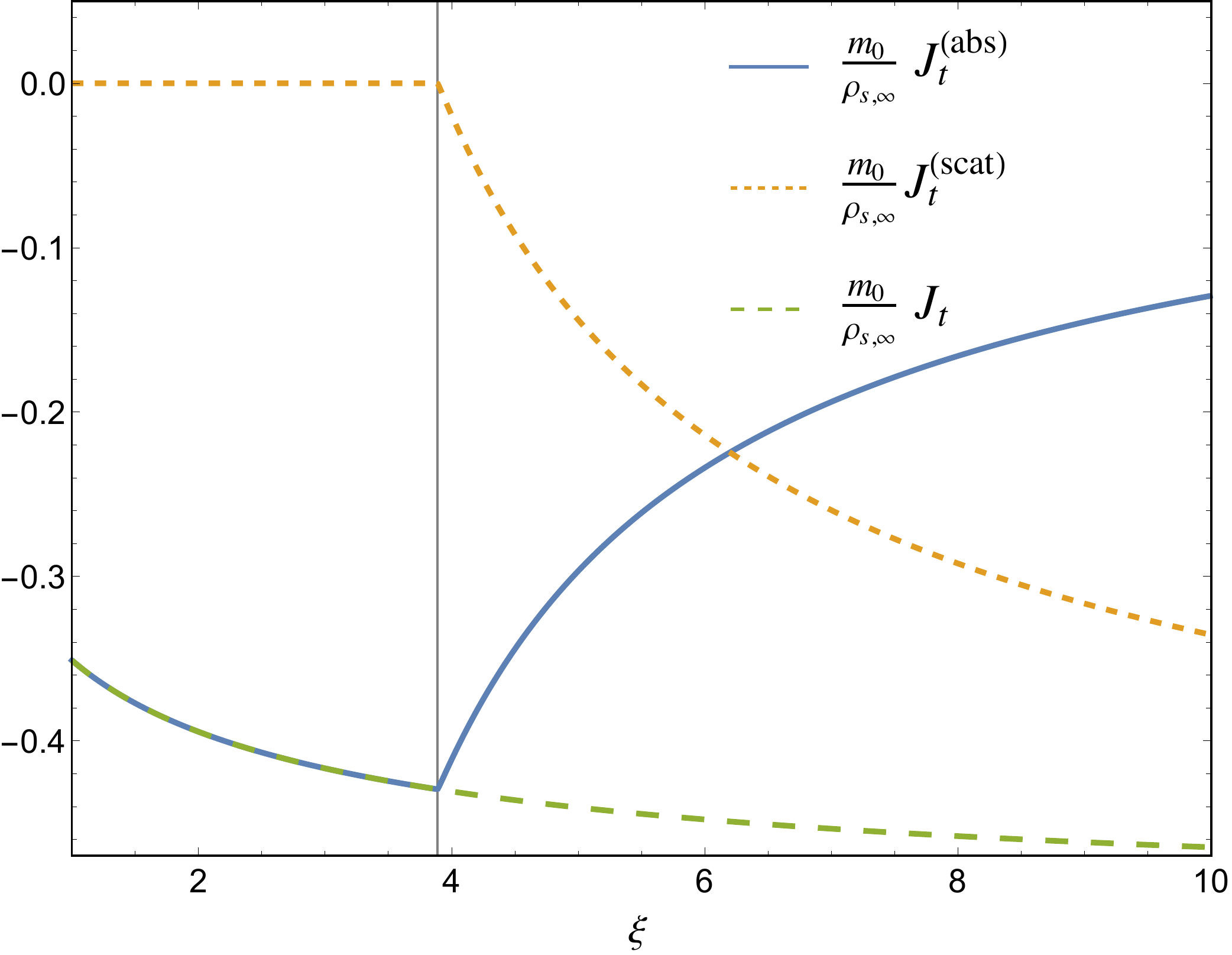}
\caption{\label{fig:jt2} Same as in Fig.\ \ref{fig:jt1}, but for $\epsilon_\sigma = -1$.}
\end{figure}

\begin{figure}
\centering
\includegraphics[width=\columnwidth]{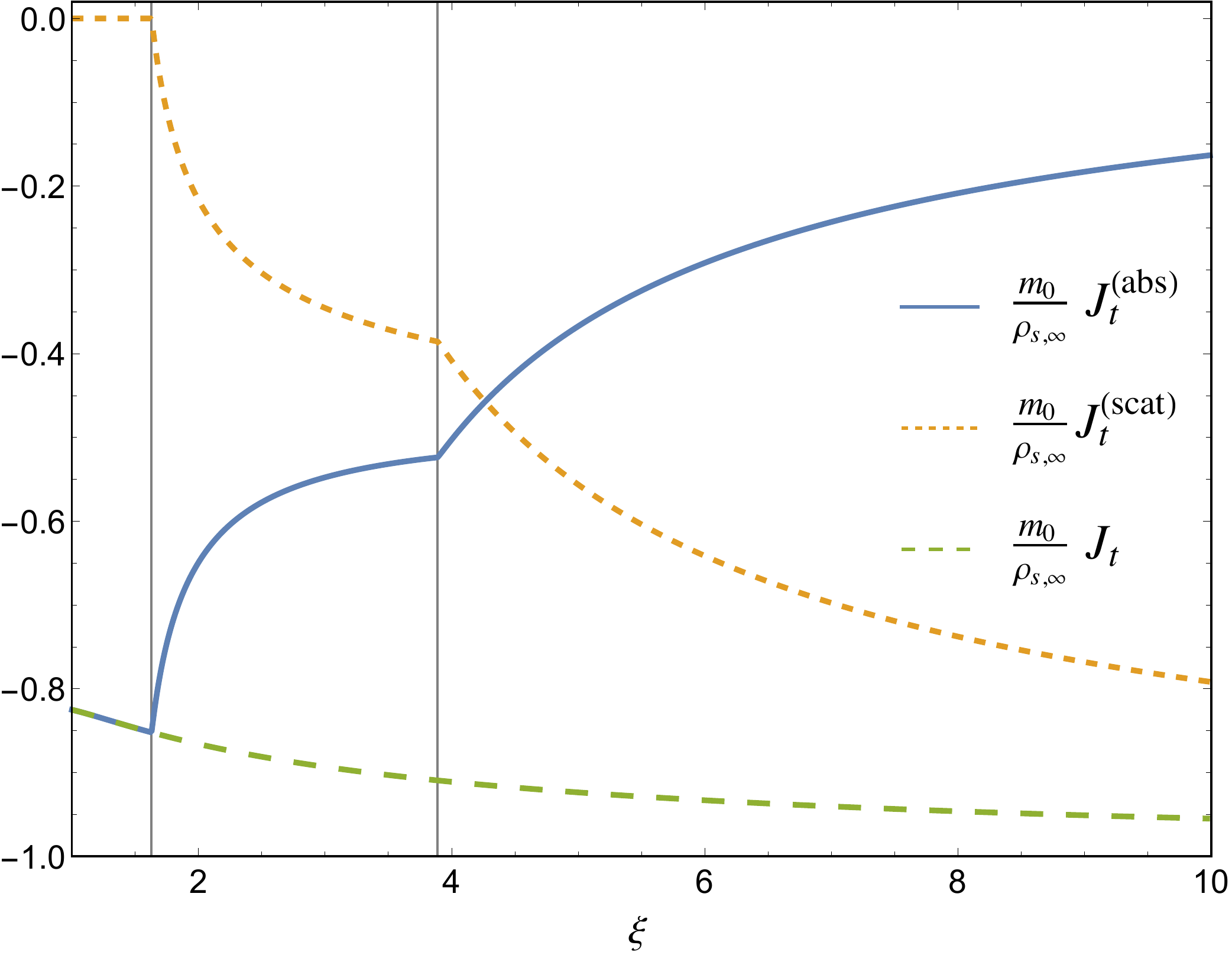}
\caption{\label{fig:jt3}
Components of the particle current surface density $J_{t}$, $J^\mathrm{(abs)}_{t}$, and $J^\mathrm{(scat)}_{t}$ obtained by summing the parts corresponding to $\epsilon_\sigma = \pm 1$. The spin parameter $\alpha = 7/8$; $\beta = 0.1$.}
\end{figure}

\begin{figure}
\centering
\includegraphics[width=\columnwidth]{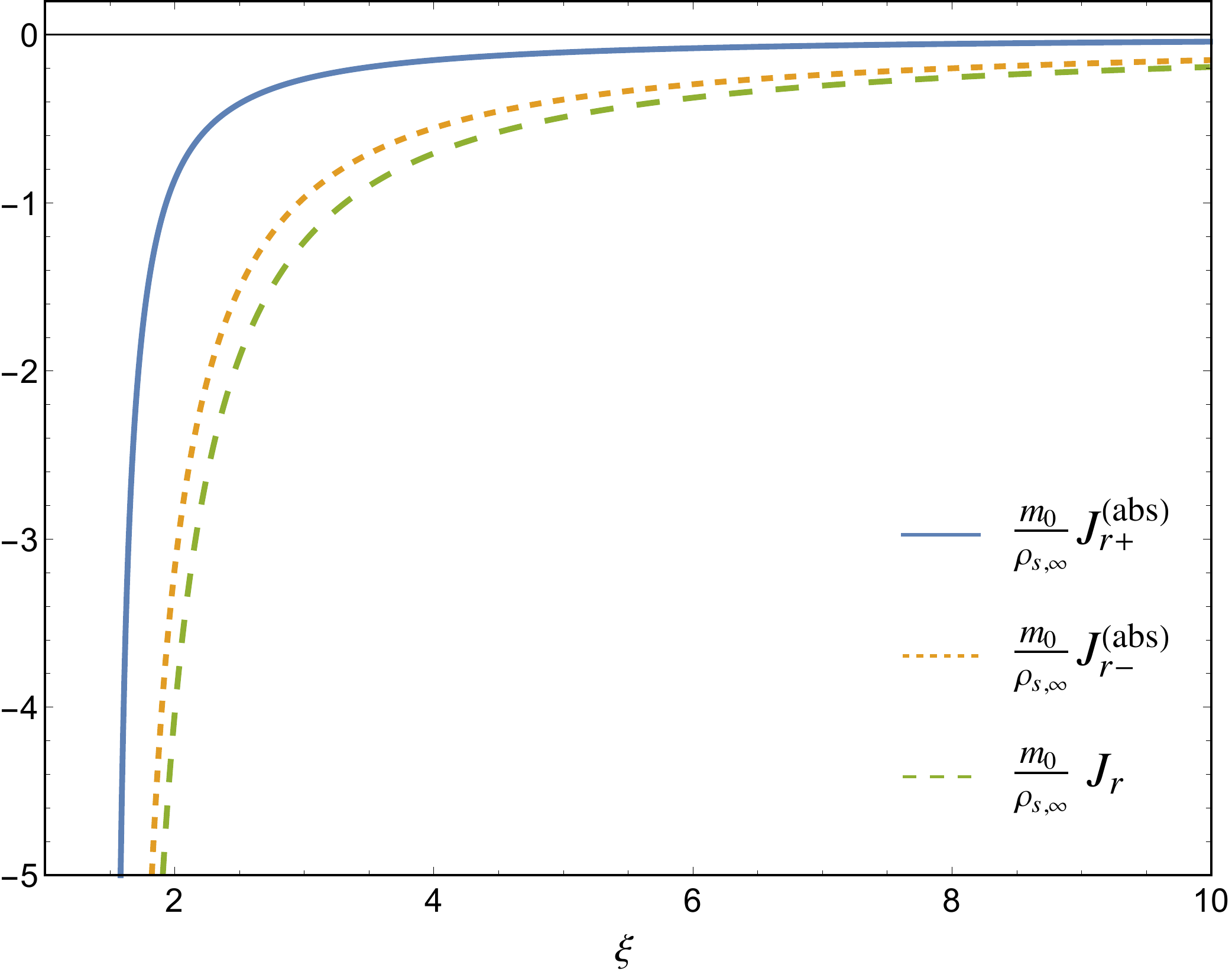}
\caption{\label{fig:jr}
Components of the particle current surface density  $J^\mathrm{(abs)}_{r+}$ (for $\epsilon_\sigma = 1$),  $J^\mathrm{(abs)}_{r-}$ (for $\epsilon_\sigma = -1$) and $J_{r}$ obtained by summing the parts corresponding to $\epsilon_\sigma = \pm 1$. The spin parameter $\alpha = 7/8$; $\beta = 0.1$.}
\end{figure}

We are interested in both absorbed and scattered unbound trajectories. By inspecting the expression \eqref{Rtilde} for $\tilde R$ at $\xi \to \infty$, we see that the unbound trajectories must satisfy $\varepsilon \ge 1$.

As in the Schwarzschild case (see, e.g., \cite{Olivier1}), a particle traveling on an absorbed trajectory cannot encounter the centrifugal barrier, when traveling from infinity towards the black hole. The centrifugal barrier is effectively controlled by the value of $\lambda$. An absorbed trajectory is characterized by the condition $\lambda \le \lambda_c(\varepsilon,\epsilon_\sigma)$, where $\lambda_c(\varepsilon,\epsilon_\sigma)$ is a solution (with respect to $\lambda$) of the set of equations
\begin{subequations}
\label{reqs}
\begin{eqnarray}
    \tilde R (\xi)  & = & 0,  \\
    \frac{d \tilde R(\xi)}{d \xi} & = & 0,
    \end{eqnarray}
\end{subequations}
and $\tilde R (\xi)$ is given by \eqref{Rtilde}. 
The function $\lambda_c(\varepsilon,\epsilon_\sigma)$ can be given in a parametric (therefore the subscript $p$) form as
\begin{subequations}
\label{parametric}
\begin{eqnarray}
\lambda_c = \lambda_p(\xi,\epsilon_\sigma) & \equiv & \frac{\xi ^{5/4}-\alpha \epsilon_\sigma  \xi ^{3/4}}{\sqrt{2 \alpha \epsilon_\sigma +(\xi -3)
   \sqrt{\xi }}}, \\
\varepsilon = \varepsilon_p(\xi,\epsilon_\sigma) & \equiv & \frac{\alpha \epsilon_\sigma +(\xi -2) \sqrt{\xi }}{\xi ^{3/4} \sqrt{2 \alpha \epsilon_\sigma +(\xi -3) \sqrt{\xi }}}.
\end{eqnarray}
\end{subequations}
It can be readily verified that the above expressions solve Eqs.\ (\ref{reqs}). Zeros of the expression
under the square root in the denominator of \eqref{parametric} correspond to circular photon orbits \eqref{xiphoton}. We have $\varepsilon_p(\xi_\mathrm{mb},\epsilon_\sigma) = 1$, where $\xi_\mathrm{mb}$ is the dimensionless radius of the marginally bound orbit, given by Eq.\ (\ref{ximb}).
On the other hand $\varepsilon_p(\xi,\epsilon_\sigma) \to \infty$ for $\xi \to \xi_\mathrm{ph}$, where $\xi_\mathrm{ph}$ is the dimensionless radius of the circular photon orbit at the equatorial plane, given by Eq.\ (\ref{xiphoton}). Consequently, the range $1 \le \varepsilon_p(\xi,\epsilon_\sigma) < \infty$ corresponds to $\xi_\mathrm{mb} \ge \xi > \xi_\mathrm{ph}$. Note that
\begin{equation}
    \frac{d \lambda_c(\varepsilon)}{d\varepsilon} = \frac{\frac{\partial \lambda_p(\xi,\epsilon_\sigma)}{\partial \xi}}{\frac{\partial \varepsilon_p(\xi,\epsilon_\sigma)}{\partial \xi}} = \xi^\frac{3}{2}.
\end{equation}

The function $\lambda_c(\varepsilon,\epsilon_\sigma)$ cannot be expressed in a closed form, however it satisfies a relatively simple equation, polynomial in both $\lambda$ and $\varepsilon$, which can be found as follows. The expressions on the left hand-side of Eqs.\ (\ref{reqs}) are polynomials in $\xi$. It is well known that the substitution $u = 1/\xi$ yields polynomials of a lower order. We define $U = \tilde R/\xi^4$, and change the variables to $u = 1/\xi$. This gives the conditions
\begin{subequations}
\label{UdU}
\begin{eqnarray}
\label{U}
    U (u)  & = &  -1 + 2 u - u^2 \alpha^2 + \varepsilon^2 - 2 \epsilon_\sigma u^2 \alpha \varepsilon \lambda \nonumber \\
    && - u^2 \lambda^2 + 2 u^3 \lambda^2 = 0, \\
    \label{dU}
    \frac{d U(u)}{du} & = & 2 - 2 u \alpha^2 - 4 \epsilon_\sigma u \alpha \varepsilon \lambda - 2 u \lambda^2 \nonumber \\
    && + 6 u^2 \lambda^2 = 0.
    \end{eqnarray}
\end{subequations}
The term $u^3$ in Eq.~\eqref{U} can be easily removed by multiplying Eq.~\eqref{dU} by $u$, solving for $u^3$, and substituting the result in Eq.~\eqref{U}. This yields the set of equations
\begin{subequations}
\begin{eqnarray}
    3 (\varepsilon^2 - 1) + 4 u - u^2 (\alpha^2 + 2 \epsilon_\sigma \alpha \varepsilon \lambda + \lambda^2) & = & 0, \\
    1 - u (\alpha^2 + 2 \epsilon_\sigma \alpha \varepsilon \lambda + \lambda^2) + 3 u^2 \lambda^2 & = & 0.
    \end{eqnarray}
\end{subequations}
instead of Eqs.\ \eqref{UdU}. Substituting temporarily
\begin{equation}
\label{subsX}
    X = \alpha^2 + 2 \epsilon_\sigma \alpha \varepsilon \lambda + \lambda^2,
\end{equation}
we get
\begin{subequations}
\label{eqsX}
\begin{eqnarray}
    3 (\varepsilon^2 - 1) + 4 u - u^2 X & = & 0, \label{3eps2}\\
    1 - u X + 3 u^2 \lambda^2 & = & 0. \label{1mux}
    \end{eqnarray}
\end{subequations}
One could continue this way in order to finally eliminate $u$, but the most straightforward method to achieve this aim is to compute the so-called Gr\"{o}bner basis for the above system. The result reads
\begin{eqnarray}
   Y  & \equiv & (\varepsilon^2 - 1) \left[ 27 \left(\varepsilon ^2-1\right) \lambda ^4-X^3+18 \lambda ^2 X \right] \nonumber \\
   && + 16 \lambda^2 - X^2 = 0.
   \label{groebner}
\end{eqnarray}
Gr\"{o}bner's basis is an invariant of the polynomial elimination, and it can be computed directly from Eqs.\ (\ref{UdU}) or \eqref{reqs}. On the other hand, to get a relatively simple form of Eq.\ (\ref{groebner}) one has to guess substitution (\ref{subsX}), suggested by Eqs.\ (\ref{eqsX}).

In practice, we compute $\lambda_c(\varepsilon,\epsilon_\sigma)$ numerically, as a solution of Eq.\ (\ref{groebner}), or we use the parametric form (\ref{parametric}). In the former case, one has to be careful to select the physical branch of solutions. Sample solution branches of Eq.\ (\ref{groebner}) are shown in Fig.\ \ref{fig:groebner}. The physical branch, parametrized by Eqs.\ (\ref{parametric}) is depicted with a red line, but other branches are also present. We show, for instance, the existence of a branch (blue) with vertical asymptote $\varepsilon = 1$ (thin line). The function $Y$ is a 6th degree polynomial in $\lambda$ and a 5th degree polynomial in $\varepsilon$. The term in $Y$ with the highest degree in $\lambda$ is $-(\varepsilon^2-1)\lambda^6$. Thus, for sufficiently large $\lambda$, $Y$ becomes positive for $\varepsilon^2 < 1$. On the other hand, Eq.\ (\ref{groebner}) has a simple solution for $\varepsilon = 1$. In this case $X = (\epsilon_\sigma \alpha + \lambda)^2$, and Eq.\ (\ref{groebner}) can be reduced to
\begin{equation}
 \label{xi4_lambda}
    Y = 16 \lambda^2 - (\epsilon_\sigma \alpha + \lambda)^4 = 0.
\end{equation}
Non-negative solutions of this equation read $\lambda = 2 - \epsilon_\sigma \alpha \pm 2 \sqrt{1 - \epsilon_\sigma \alpha}$,
and the larger of the two solutions is
\begin{equation}
    \lambda_\ast = 2 - \epsilon_\sigma \alpha + 2 \sqrt{1 - \epsilon_\sigma \alpha}
\end{equation}
(note an algebraic coincidence $\lambda_\ast = \xi_\mathrm{mb}$). The remaining two non-positive solutions are $\lambda = -2 - \epsilon_\sigma \alpha \pm 2 \sqrt{1 + \epsilon_\sigma \alpha}$. It follows that for $\varepsilon=1$ and $\lambda > \lambda_\ast$, we have $Y < 0$. Consequently, Eq.\ (\ref{groebner}) has an asymptotic solution with $\lambda \to +\infty$ for $\varepsilon \to 1$. 

Equation (\ref{groebner}) has a simple solution for $\alpha = \pm 1$ and $\epsilon_\sigma \alpha = + 1$, i.e., for prograde orbits in the extremal Kerr metric. In this case $\lambda_c(\varepsilon,\epsilon_\sigma) = \varepsilon$. This solution is shown in Fig.\ \ref{fig:groebner3}.

For scattered orbits, $\lambda_c(\varepsilon,\epsilon_\sigma) < \lambda \le \lambda_\mathrm{max}(\xi,\varepsilon,\epsilon_\sigma)$. The formula for the upper limit $\lambda_\mathrm{max}(\xi,\varepsilon,\epsilon_\sigma)$ follows from the requirement that $\tilde R \ge 0$. It reads
\begin{widetext}
\begin{equation}
\lambda_\mathrm{max}(\xi,\varepsilon,\epsilon_\sigma) = \frac{\xi}{\xi - 2} \left\{ \sqrt{\alpha^2 \left( \varepsilon^2 + \frac{2}{\xi} - 1 \right)+(\xi - 2) \left[\xi  \left(\varepsilon^2 - 1 \right) + 2\right]} - \epsilon_\sigma \alpha \varepsilon \right\}.
\end{equation}
\end{widetext}
Note that for $|\alpha| \ge \sqrt{2}/2$ we have $\xi_\text{ph} \le 2$ and $\lambda_\mathrm{max}$ must be evaluated at the removable discontinuity $\xi=2$ for prograde trajectories. The relevant limit for $\epsilon_\sigma \alpha  >0$ is
\begin{equation}
\lim_{\xi \to 2} \lambda_\text{max}(\xi, \varepsilon, \epsilon_\sigma) = \frac{4 \varepsilon^2 - \alpha^2}{2 \epsilon_\sigma \alpha  \varepsilon}.
\end{equation}

Scattered unbound orbits exist only for sufficiently high energies $\varepsilon$. The main bound, $\varepsilon > 1$, applies to all trajectories reaching infinity. On the other hand, a condition defining a scattered orbit is the existence of a centrifugal barrier---a particle on a scattered orbit travels from infinity, reaches a turning point, at which $\tilde R(\xi) = 0$, and moves again to infinity. In other words, a scattered orbit parametrized by $\varepsilon$, $\lambda$, and $\epsilon_\sigma$ can only pass through a point of radius $\xi$, if there is a turning point at $\xi_0 \le \xi$ (i.e., $\tilde R(\xi_0) = 0$) and the region with radii $\xi^\prime \ge \xi$ is available for motion (i.e., $\tilde R(\xi^\prime) \ge 0$). In order to obtain the value of the minimal energy $\varepsilon_\mathrm{min}(\xi,\epsilon_\sigma)$ allowing for a scattered orbit at some radius $\xi$, we solve Eqs.\ (\ref{eqsX}) with respect to $\varepsilon$. More precisely, we start by solving Eq.\ (\ref{1mux}) with respect to $\lambda$. There are two possible solutions, which we can substitute in Eq.\ (\ref{3eps2}). Solving the resulting equations with respect to $\varepsilon$ yields 4 solutions, two of which can be positive. These are
\begin{widetext}
\begin{equation}
\label{eminaux}
    \varepsilon = \sqrt{\frac{\pm 2 \sqrt{\frac{\left[\alpha ^3+\alpha  (\xi -2) \xi \right]^2}{\xi }}+\alpha ^2 (5-3 \xi )+(\xi -3) (\xi -2)^2 \xi }{\xi 
   \left[(\xi -3)^2 \xi -4 \alpha ^2\right]}};
\end{equation}
they do not depend explicitly on $\epsilon_\sigma$, but in fact one of them corresponds to prograde, and the other to retrograde orbits. One can distinguish between the two cases by inspecting the asymptotics of the two expressions. The two solutions tend to infinity for
\begin{equation}
    \xi \to 2 + 2 \cos \left[ \frac{2}{3} \mathrm{arccos}( \pm \alpha) \right],
\end{equation}
i.e., solutions to $(\xi - 3)^2 \xi - 4 \alpha^2 = 0$. These solutions can be easily identified as the locations of circular photon orbits at the equatorial plane. The latter are given by $\xi = \xi_\mathrm{ph}$, where $\xi_\mathrm{ph}$ is defined by Eq.\ (\ref{xiphoton}).

Matching the asymptotics of Eq.\ (\ref{eminaux}) with the locations of circular photon orbits, we can select the signs in (\ref{eminaux}), leading to the formula
\begin{equation}
\varepsilon = \sqrt{\frac{- 2 \epsilon_\sigma \alpha \left[\alpha^2 + (\xi -2) \xi \right] \xi^{-1/2}+\alpha ^2 (5-3 \xi )+(\xi -3) (\xi -2)^2 \xi }{\xi  \left[(\xi -3)^2 \xi -4
   \alpha ^2\right]}}.
\end{equation}
For both solutions (with $\epsilon_\sigma = \pm 1$) the limiting energy given by the above formula decreases from $+\infty$ for $\xi > \xi_\mathrm{ph}$. The value $\varepsilon = 1$ is reached at $\xi = \xi_\mathrm{mb}$, where $\xi_\mathrm{mb}$ is given by Eq.\ (\ref{ximb}). Thus, the minimal allowed energy for scattered orbits can be written as
\begin{equation}
    \varepsilon_\mathrm{min}(\xi,\epsilon_\sigma) = \begin{cases} \infty & \text{for} \quad \xi < \xi_\mathrm{ph}, \\
    \sqrt{\frac{- 2 \epsilon_\sigma \alpha \left[\alpha ^2 +  (\xi -2) \xi \right] \xi^{-1/2}+\alpha ^2 (5-3 \xi )+(\xi -3) (\xi -2)^2 \xi }{\xi  \left[(\xi -3)^2 \xi -4
   \alpha ^2\right)]}} & \text{for} \quad \xi_\mathrm{ph} < \xi < \xi_\mathrm{mb}, \\
   1 & \text{for} \quad \xi \ge \xi_\mathrm{mb}.
    \end{cases}
\end{equation}
\end{widetext}
Note that no scattered trajectory can extend to $\xi < \xi_\mathrm{ph}$. Sample graphs of $\varepsilon_\mathrm{min}(\xi,\epsilon_\sigma)$ are shown in Fig.\ \ref{fig:emin}.

\section{Particle current surface density}
\label{sec:current}

We divide the expressions for the particle current surface density $J_\mu$ into two parts: a part corresponding to absorbed trajectories and a part corresponding to scattered ones. The quantities referring to those two parts will be denoted with the superscripts (abs) and (scat), respectively. Consequently, $J_\mu = J_\mu^\mathrm{(abs)} + J_\mu^\mathrm{(scat)}$. As discussed in the previous section, the phase-space region corresponding to absorbed trajectories is characterized by $\varepsilon \ge 1$  and $\lambda \le \lambda_c(\varepsilon,\epsilon_\sigma)$. Scattered orbits passing through a point at radius $\xi$ are characterized by $\varepsilon_\mathrm{min}(\xi,\epsilon_\sigma) < \varepsilon < \infty$ and $\lambda_c(\varepsilon,\epsilon_\sigma) < \lambda \le \lambda_\mathrm{max}(\xi,\varepsilon,\epsilon_\sigma)$. Explicit expressions for $J_\mu^\mathrm{(abs)}$ and $J_\mu^\mathrm{(scat)}$ read
\begin{widetext}
\begin{subequations}
\begin{eqnarray}
J^\mathrm{(abs)}_t (\xi) & = & - A m_0^3 \xi \sum_{\epsilon_\sigma = \pm 1} \int_1^\infty d \varepsilon \exp(- \beta \varepsilon) \varepsilon \int_0^{\lambda_c(\varepsilon,\epsilon_\sigma)} \frac{d \lambda}{\sqrt{\tilde R}}, \\
J^\mathrm{(abs)}_r (\xi) & = & - \frac{A M^2 m_0^3 \xi}{\Delta} \sum_{\epsilon_\sigma = \pm 1} \int_1^\infty d \varepsilon \exp(- \beta \varepsilon)  \lambda_c(\varepsilon,\epsilon_\sigma), \\
J^\mathrm{(abs)}_\varphi (\xi) & = & A M m_0^3 \xi \sum_{\epsilon_\sigma = \pm 1} \int_1^\infty d \varepsilon \exp(- \beta \varepsilon) \int_0^{\lambda_c(\varepsilon,\epsilon_\sigma)} d \lambda  \frac{\epsilon_\sigma \lambda + \alpha \varepsilon}{\sqrt{\tilde R}}
\end{eqnarray}
\end{subequations}
and
\begin{subequations}
\begin{eqnarray}
J^\mathrm{(scat)}_t (\xi) & = & - 2 A m_0^3 \xi \sum_{\epsilon_\sigma = \pm 1} \int_{\varepsilon_\mathrm{min}(\xi,\epsilon_\sigma)}^\infty d \varepsilon \exp(- \beta \varepsilon) \varepsilon  \int_{\lambda_c(\varepsilon,\epsilon_\sigma)}^{\lambda_\mathrm{max}(\xi,\varepsilon,\epsilon_\sigma)} \frac{d\lambda}{\sqrt{\tilde R}}, \label{jtscat} \\
J^\mathrm{(scat)}_r (\xi) & = & 0, \\
J^\mathrm{(scat)}_\varphi (\xi) & = & 2 A M m_0^3 \xi \sum_{\epsilon_\sigma = \pm 1} \int_{\varepsilon_\mathrm{min}(\xi,\epsilon_\sigma)}^\infty d \varepsilon \exp(- \beta \varepsilon) \int_{\lambda_c(\varepsilon,\epsilon_\sigma)}^{\lambda_\mathrm{max}(\xi,\varepsilon,\epsilon_\sigma)} d \lambda \frac{\epsilon_\sigma \lambda + \alpha \varepsilon}{\sqrt{\tilde R}}. \label{jphiscat}
\end{eqnarray}
\end{subequations}
The factor 2 in Eqs.\ (\ref{jtscat}) and (\ref{jphiscat}) is due to two possible radial directions of motion along a scattered trajectory, $\epsilon_r = \pm 1$. Note that in dimensionless variables $\Delta = M^2 (\xi^2-2 \xi  +\alpha^2)$.

Using the parametric solution for $\lambda_c(\varepsilon,\epsilon_\sigma)$ given by Eq.\ (\ref{parametric}), we get
\begin{subequations}
\begin{eqnarray}
J^\mathrm{(abs)}_t (\xi) & = & - A m_0^3 \xi \sum_{\epsilon_\sigma = \pm 1} \int_{\xi_\mathrm{mb}}^{\xi_\mathrm{ph}} d \bar \xi \varepsilon_p^\prime (\bar \xi, \epsilon_\sigma) \varepsilon_p(\bar \xi, \epsilon_\sigma) \exp[- \beta \varepsilon_p(\bar \xi, \epsilon_\sigma)] \int_0^{\lambda_p(\bar \xi,\epsilon_\sigma)} \frac{d \lambda}{\sqrt{\tilde R[\varepsilon_p(\bar \xi, \epsilon_\sigma),\lambda,\xi,\epsilon_\sigma]}}, \\
J^\mathrm{(abs)}_r (\xi) & = & - \frac{A M^2 m_0^3 \xi}{\Delta} \sum_{\epsilon_\sigma = \pm 1} \int_{\xi_\mathrm{mb}}^{\xi_\mathrm{ph}} d \bar \xi \varepsilon_p^\prime(\bar \xi, \epsilon_\sigma) \exp[- \beta \varepsilon_p(\bar \xi, \epsilon_\sigma)]  \lambda_p(\bar \xi,\epsilon_\sigma), \\
J^\mathrm{(abs)}_\varphi (\xi) & = & A M m_0^3 \xi \sum_{\epsilon_\sigma = \pm 1} \int_{\xi_\mathrm{mb}}^{\xi_\mathrm{ph}} d \bar \xi \varepsilon_p^\prime (\bar \xi,\epsilon_p) \exp[- \beta \varepsilon_p(\bar \xi,\epsilon_\sigma)] \int_0^{\lambda_p(\bar \xi,\epsilon_\sigma)} d \lambda  \frac{\epsilon_\sigma \lambda + \alpha \varepsilon_p(\bar \xi,\epsilon_\sigma)}{\sqrt{\tilde R[\varepsilon_p(\bar \xi,\epsilon_\sigma),\lambda,\xi,\epsilon_\sigma]}},
\end{eqnarray}
\end{subequations}
where $\varepsilon_p^\prime (\xi,\epsilon_\sigma) = \partial \varepsilon_p (\xi,\epsilon_p)/\partial \xi$, and where we have explicitly listed the arguments in $\tilde R$, i.e., we set
\begin{equation}
    \tilde R(\varepsilon,\lambda,\xi,\epsilon_\sigma) = \xi^4 \left[ \varepsilon^2 - \left( 1 - \frac{2}{\xi} \right) \left( 1 + \frac{\lambda^2}{\xi^2} \right) - \frac{2 \epsilon_\sigma \alpha \varepsilon \lambda + \alpha^2}{\xi^2} \right],
\end{equation}
\end{widetext}
to avoid confusion. In principle, a similar parametrization can be used to express the integrals corresponding to scattered trajectories, however it is difficult to find the integration limit corresponding to $\varepsilon_\mathrm{min}(\xi,\epsilon_\sigma)$, i.e., $\bar \xi$ such that $\varepsilon_\mathrm{min}(\xi,\epsilon_\sigma) = \varepsilon_p(\bar \xi,\epsilon_\sigma)$.

Sample graphs of various components of the particle current surface density are shown in Figs.\ \ref{fig:jmu1}--\ref{fig:jr}. The main feature visible in these plots is the cutoff for the scattered components $J_t^\mathrm{(scat)}$, $J_r^\mathrm{(scat)}$, and $J_\varphi^\mathrm{(scat)}$ at the location of the circular photon orbit $\xi_\mathrm{ph}$, which in turn depends on $\epsilon_\sigma$: $\epsilon_\sigma=+1$ (prograde motion, Fig.~\ref{fig:jmu1}); $\epsilon_\sigma=-1$ (retrograde motion, Fig.~\ref{fig:jmu2}). The difference is best visible in Fig.~\ref{fig:jt3}. Note that the graphs of the components $J_t^\mathrm{(abs)}$, $J_t^\mathrm{(scat)}$, $J_\varphi^\mathrm{(abs)}$, and $J_\varphi^\mathrm{(scat)}$ are not smooth, due to the cutoff behavior at $\xi = \xi_\mathrm{ph}$.

\begin{figure}
    \centering
    \includegraphics[width=\columnwidth]{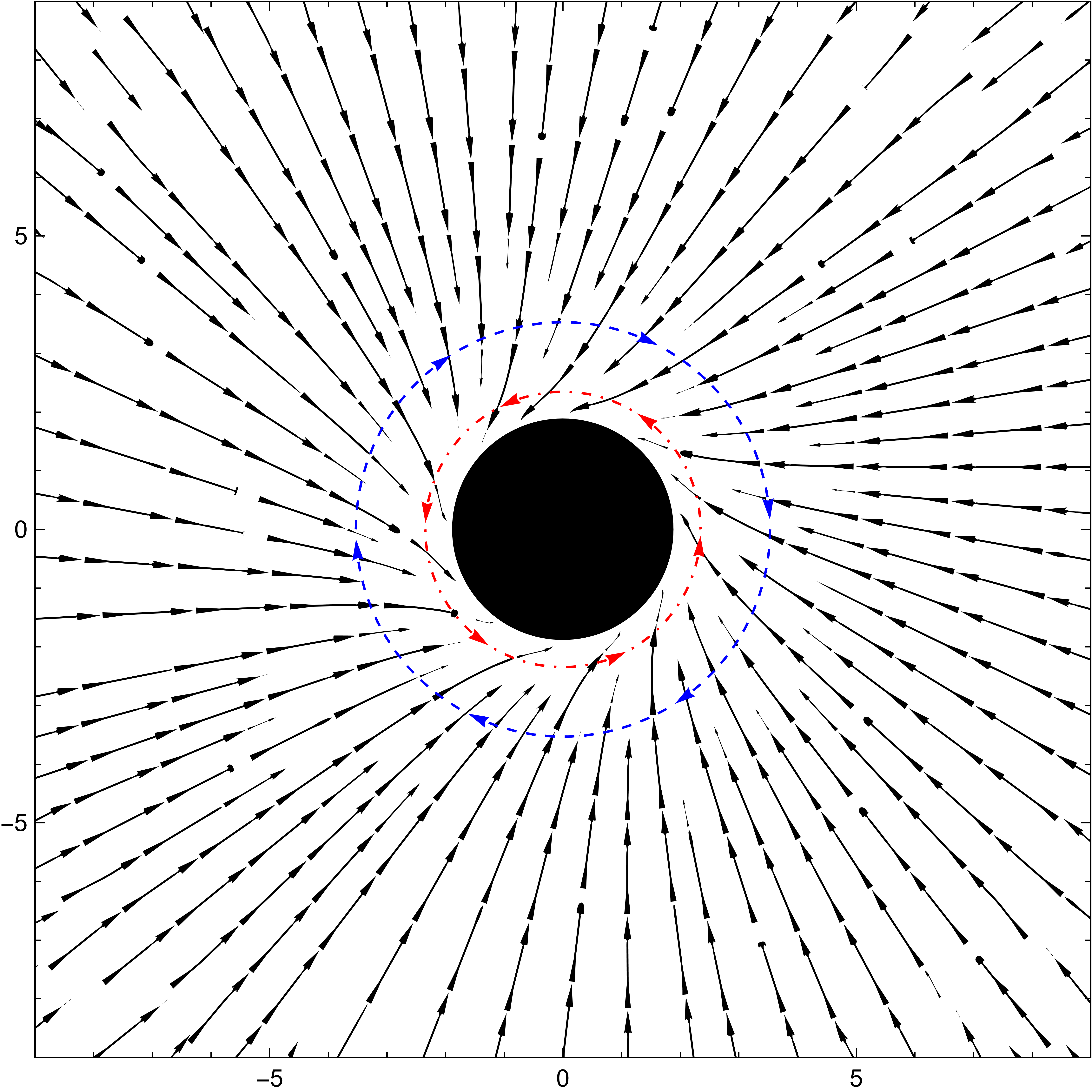}
    \caption{\label{fig:streamplot} Streamlines of the vector field $(J^x, J^y)$. Circular photon orbits are are shown in red and blue colors. The region inside the black hole horizon is marked in black. The black hole rotates counterclockwise. The parameters used to create this plot are $\alpha = 0.5$ and $\beta = 1$. The units on the axes are mass units $M$.}
\end{figure}

Figure \ref{fig:streamplot} shows the directions of the flow at the equatorial plane for $\alpha = 1/2$ and $\beta = 1$. We use standard Cartesian coordinates defined by $t^\prime = t$, $x =  r \cos \varphi$, $y = r \sin \varphi$. This yields the following contravariant components of the particle current surface density:
\begin{subequations}
\begin{eqnarray}
J^{t'} & = & -\frac{J_t \left[\xi^3 + (\xi +2) \alpha ^2 \right]+2 \alpha  J_\varphi/M }{  (\xi^2 -2 \xi + \alpha^2) \xi }, \\
J^x & = & J_r  \frac{\xi^2 -2 \xi + \alpha^2}{\xi^2} \cos{\varphi} \nonumber \\
&& - \frac{J_\varphi/M \times (\xi-2)-2 \alpha J_t }{\xi^2 -2 \xi + \alpha^2 } \sin{\varphi}, \\
J^y & = & J_r \frac{\xi^2 -2 \xi + \alpha^2}{\xi^2} \sin{\varphi} \nonumber \\
&& + \frac{ J_\varphi/M \times (\xi-2) - 2 \alpha J_t  }{\xi^2 -2 \xi + \alpha^2 } \cos{\varphi}.
\end{eqnarray}
\end{subequations}
The arrows in Fig.\ \ref{fig:streamplot} correspond to the components $(J^x,J^y)$. The accretion flow is nearly radial outside the outer (retrograde) circular photon orbit; it twists towards the prograde circular photon orbit, where the dragging effect becomes much more pronounced. Finally, the matter sinks below the black hole horizon  

\begin{figure*}
    \centering
    \includegraphics[width=\columnwidth]{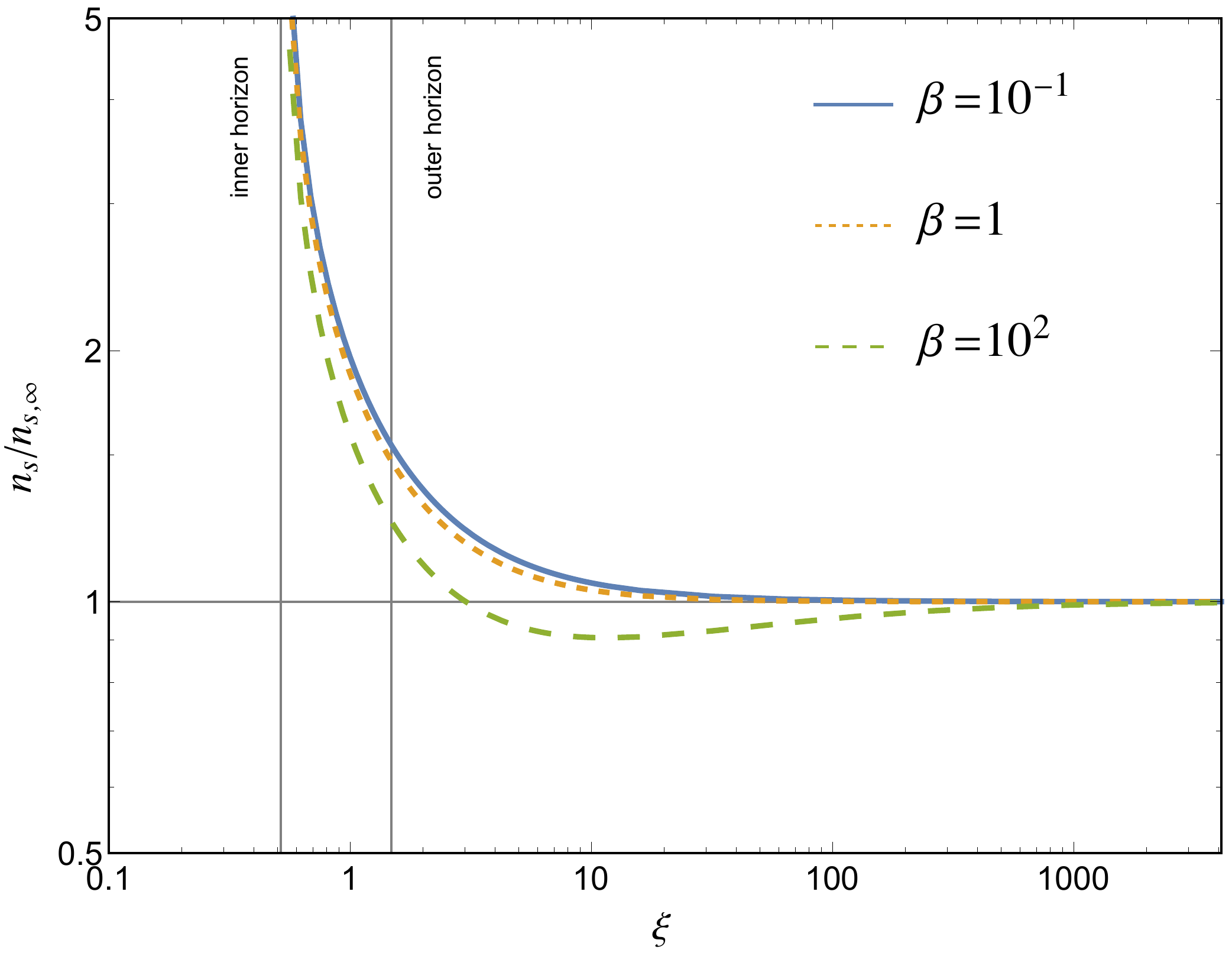}~\includegraphics[width=\columnwidth]{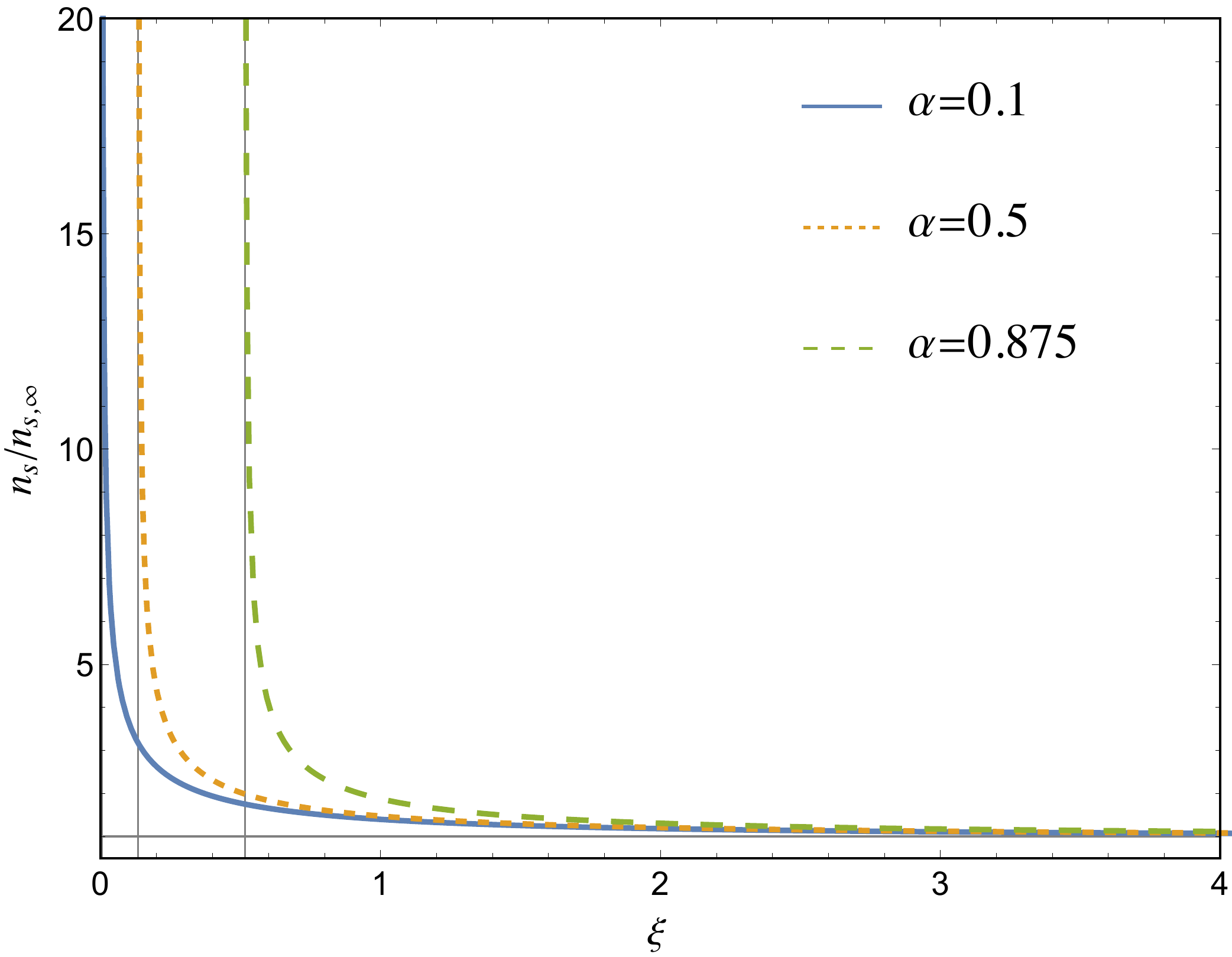}
    \caption{\label{fig:n} The invariant particle surface density $n_s$ vs.\ the dimensionless radius $\xi$. The left panel shows the dependence of $n_s$ on the value of the parameter $\beta$ for $\alpha = 7/8$. In the right panel we set $\beta = 1$ and plot the graphs of $n_s$ for three different values of $\alpha$. Vertical lines in the right plot correspond to inner Kerr horizons.}
\end{figure*}

We should emphasize that the vector components $J_\mu$ depend on the assumed metric gauge, in particular on the chosen time foliation. This is best visible for the component $J_r$ (cf. Fig.\ \ref{fig:jr}), which in the Boyer-Lindquist coordinates is divergent at the black hole horizon. On the other hand, the particle surface density $n_s$ defined by Eq.\ \eqref{n_s} and plotted in Fig.\ \ref{fig:n} is a gauge invariant quantity. The particle surface density $n_s$ is continuous across the outer horizon, but it diverges at the inner Kerr horizon (Fig.\ \ref{fig:n}, right panel). This latter behavior resembles the so-called mass inflation discovered by Poisson and Israel \cite{poisson}. A peculiar feature illustrated in Fig.\ \ref{fig:n} is a local minimum of the particle surface density $n_s$ near the retrograde marginally stable orbit, occurring for high values of the parameter $\beta$. This behavior seems to be related with the dimensionality of the model---it has not been observed for spherically symmetric models investigated in \cite{Olivier1,Olivier2,cieslik}.

\section{Accretion rates}
\label{sec:accretionrates}

The two Killing vectors admitted by the metric $\gamma$ induced at the equatorial plane (\ref{kerr3d}), i.e., $\xi^\mu = (\xi^t,\xi^r,\xi^\varphi) = (1,0,0)$ and $\chi^\mu = (\chi^t,\chi^r,\chi^\varphi) = (0,0,1)$, give rise to two independent conserved currents $J^\mu_{(t)} = T\indices{^\mu_\nu} \xi^\nu$ and $J^\mu_{(\varphi)} =  T\indices{^\mu_\nu} \chi^\nu$, satisfying $\nabla_\mu J^\mu_{(t)} = 0$ and $\nabla_\mu J^\mu_{(\varphi)} = 0$. Together with the conserved particle current surface density $J^\mu$, they yield three accretion rates, which can be defined basing on the Stokes theorem.

Let $\mathcal M$ be a region in a $d$ dimensional spacetime equipped with metric ${}^{(d)}g$. Let $\partial \mathcal M$ denote the $d-1$ dimensional boundary of $\mathcal M$, and let ${}^{(d-1)}g$ be the metric induced on $\partial \mathcal M$. A general-relativistic version of the Stokes theorem can be written as
\begin{eqnarray}
   \lefteqn{ \int_{\mathcal M} d^d x \sqrt{|\mathrm{det} \, {}^{(d)}g|} \nabla_\mu V^\mu } \nonumber \\
   && = \int_{\partial \mathcal M} d^{d-1}y \sqrt{|\mathrm{det} \, {}^{(d-1)}g|} n_\mu V^\mu.
\end{eqnarray}
Here the covariant derivative $\nabla_\mu$ is defined with respect to the metric ${}^{(d)}g$, $V^\mu$ is a vector field on $\mathcal M$, and $n^\mu$ is the vector field normal to the boundary $\partial \mathcal M$. If $\nabla_\mu V^\mu = 0$, we get
\begin{equation}
\label{surfaceterm}
    \int_{\partial \mathcal M} d^{d-1} y \sqrt{|\mathrm{det} \, {}^{(d-1)}g|} n_\mu V^\mu = 0.
\end{equation}

Working in the 2+1 dimensional case restricted to the equatorial plane only, we take $d = 3$ and ${}^{(d)}g = {}^{(3)}g = \gamma$. Assume $\mathcal M$ to be a region enclosed by two surfaces of constant time $\Sigma_1$, $\Sigma_2$ and restricted by $r_1 < r < r_2$. The metric induced on the surface $r = \mathrm{const}$, $\theta = \pi/2$ reads
\begin{eqnarray}
    ^{(2)}g & = & \tilde \gamma = \left( -1 + \frac{2 M}{r}\right) dt^2 - \frac{4 M a}{r} dt d \varphi \nonumber \\
    && + \left[ r^2 + a^2 \left( 1 + \frac{2M}{r} \right) \right] d \varphi^2.
\end{eqnarray}
We have $\mathrm{det} \, \tilde \gamma_{\mu \nu} = - \Delta$. The vector normal to the surface $r = \mathrm{const}$ has the components $\tilde n_r = \pm r /\sqrt{\Delta}$, $\tilde n_t = \tilde n_\varphi = 0$.
The flux through the boundary at $r = r_1$ can be written as
\begin{equation}
    \int_{r = r_1} d^2 y \sqrt{|\mathrm{det} \, \tilde \gamma|} \tilde n_\mu V^\mu = - \int_{r = r_1} dt d\varphi r V^r,
\end{equation}
giving rise to the accretion rate
\begin{equation}
    \dot V = - \int_{r=r_1} r V^r d \varphi.
\end{equation}
The minus sign in the above definition is a matter of convention, but it has a meaning. Here, and in what follows, we choose the signs so that the accretion rates correspond to the fluxes into the black hole.

Assuming $V^\mu = m_0 J^\mu$, $- J^\mu_{(t)}$, or $J^\mu_{(\varphi)}$, we get 3 accretion rates, respectively: the rest-mass accretion rate given by $\dot M = - 2 \pi M m_0 \xi J^r$, the energy accretion rate $\dot{\mathcal E} = 2 \pi M \xi T\indices{^r_t}$, and the angular momentum accretion rate given by $\dot{\mathcal L} = -2 \pi M \xi T\indices{^r_\varphi}$. Sign conventions in the definitions of $\dot{\mathcal{E}}$ and $\dot{\mathcal L}$ can be checked by noticing that for perfect fluids $T\indices{^r_t} \propto u^r u_t$ and $T\indices{^r_\varphi} \propto u^r u_\varphi$, where $u^\mu$ denote the components of the four-velocity of the fluid.

The expression for $J^r$ given by Eq.\ (\ref{jrupgeneral}) yields the rest-mass accretion rate
\begin{eqnarray}
    \dot M & = & - 2 \pi M m_0 \xi J^r \nonumber \\
    & = & - 2 \pi A M m_0^4 \int \epsilon_r \exp ( - \beta \varepsilon ) d \varepsilon d \lambda.
\end{eqnarray}
Since $J_r^\mathrm{(scat)} = 0$, we only get a contribution from absorbed trajectories with $1 \le \varepsilon$ and $\lambda \le \lambda_c(\varepsilon,\epsilon_\sigma)$, i.e, 
\begin{equation}
\label{dotM_nonparam}
    \dot M =  2 \pi A M m_0^4 \sum_{\epsilon_\sigma = \pm 1} \int_1^\infty d \varepsilon \exp ( - \beta \varepsilon ) \lambda_c(\varepsilon,\epsilon_\sigma).
\end{equation}

\begin{widetext}
For the energy and the angular momentum accretion rates we get
\begin{eqnarray}
\label{dotE_nonparam}
    \dot{\mathcal{E}} & = & 2 \pi M \xi T\indices{^r_t} = 2 \pi A m_0^4 M \sum_{\epsilon_\sigma = \pm 1} \int_1^\infty d \varepsilon \exp(- \beta \varepsilon) \varepsilon \lambda_c(\varepsilon,\epsilon_\sigma), \\
\label{dotL_nonparam}
    \dot{\mathcal{L}} & = & - 2 \pi M \xi T\indices{^r_\varphi} = 2 \pi A m_0^4 M^2 \sum_{\epsilon_\sigma = \pm 1} \int_1^\infty d \varepsilon \exp(- \beta \varepsilon) \left[ \frac{\epsilon_\sigma}{2} \lambda_c^2(\varepsilon,\epsilon_\sigma) + \alpha \varepsilon \lambda_c(\varepsilon,\epsilon_\sigma) \right],
\end{eqnarray}
respectively. Note that $\dot{\mathcal L} = 0$ for a non-rotating black hole, i.e., for $\alpha = 0$, since in this case $\lambda_c(\varepsilon,+1) = \lambda_c(\varepsilon,-1)$.

Finally, expressing the normalization constant $A$ in terms of the asymptotic surface rest-mass density $\rho_{s,\infty}$ \eqref{A} or the asymptotic surface energy density $\varepsilon_{s,\infty} = - \lim_{\xi \to \infty} T\indices{^t_t}$, we get
\begin{subequations}
\label{ratesepsilon}
\begin{eqnarray}
\label{mdotenergy}
    \dot M & = &  M \rho_{s,\infty} \frac{\beta^2 \exp(\beta)}{1 + \beta}  \sum_{\epsilon_\sigma = \pm 1} \int_1^\infty d \varepsilon \exp ( - \beta \varepsilon ) \lambda_c(\varepsilon,\epsilon_\sigma), \\
    \dot{\mathcal{E}} & = & M \varepsilon_{s,\infty} \frac{\beta^3 \exp(\beta)}{2 + 2 \beta + \beta^2}  \sum_{\epsilon_\sigma = \pm 1} \int_1^\infty d \varepsilon \exp(- \beta \varepsilon) \varepsilon \lambda_c(\varepsilon,\epsilon_\sigma), \\
    \dot{\mathcal{L}} & = & M^2 \varepsilon_{s,\infty} \frac{\beta^3 \exp(\beta)}{2 + 2 \beta + \beta^2}  \sum_{\epsilon_\sigma = \pm 1} \int_1^\infty d \varepsilon \exp(- \beta \varepsilon) \left[ \frac{\epsilon_\sigma}{2} \lambda_c^2(\varepsilon,\epsilon_\sigma) + \alpha \varepsilon \lambda_c(\varepsilon,\epsilon_\sigma) \right].
\end{eqnarray}
\end{subequations}
The results of the remainder of this section suggest that the parametrization in terms of the asymptotic surface energy density $\varepsilon_{s,\infty}$ seems to be more relevant to accretion rates $\dot{\mathcal{E}}$ and $\dot{\mathcal{L}}$ than an alternative one which uses $\rho_{s,\infty}$. The two quantities are related by
\begin{equation}
    \rho_{s,\infty} = \frac{\beta (1 + \beta)}{2 + 2 \beta + \beta^2} \varepsilon_{s,\infty}.
\end{equation}
We derive this relation in Appendix \ref{sec:minkowski}.

The expressions for $\dot M$, $\dot{\mathcal{E}}$, $\dot{\mathcal{L}}$, understood as functions of $\beta$, are given by Laplace transforms of certain functions. For instance, $\dot M$ turns out to be a Laplace transform of $\lambda_c$. This Laplace form implies certain standard relations, such as
\begin{equation}
    - \frac{\partial}{\partial \beta} \int_1^\infty d \varepsilon \exp(- \beta \varepsilon) \lambda_c(\varepsilon,\epsilon_\sigma) = \int_1^\infty d \varepsilon \exp(- \beta \varepsilon) \varepsilon \lambda_c(\varepsilon,\epsilon_\sigma).
\end{equation}
Thus $\dot M$ and $\dot{\mathcal{E}}$ are related.

Using the parametric representation of $\lambda_c(\varepsilon)$ given by Eq.\ (\ref{parametric}), one can express the accretion rates as
\begin{subequations}
\label{dotMEL_param}
\begin{eqnarray}
\label{dotM_param}
\dot M & = & - M \rho_{s,\infty} \frac{\beta^2 \exp(\beta)}{1 + \beta}  \sum_{\epsilon_\sigma = \pm 1} \int_{\xi_\mathrm{ph}}^{\xi_\mathrm{mb}} d \bar \xi \varepsilon_p^\prime (\bar \xi,\epsilon_\sigma) \exp [ - \beta \varepsilon_p(\bar \xi,\epsilon_\sigma) ] \lambda_p(\bar \xi,\epsilon_\sigma), \\
\label{dotE_param}
\dot{\mathcal E} & = & - M \varepsilon_{s,\infty} \frac{\beta^3 \exp(\beta)}{2 + 2 \beta + \beta^2} \sum_{\epsilon_\sigma = \pm 1} \int^{\xi_\mathrm{mb}}_{\xi_\mathrm{ph}} d \bar \xi \varepsilon_p^\prime(\bar \xi,\epsilon_\sigma) \exp[- \beta \varepsilon_p(\bar \xi,\epsilon_\sigma)] \varepsilon_p(\bar \xi,\epsilon_\sigma) \lambda_p(\bar \xi,\epsilon_\sigma),\\
\label{dotL_param}
\dot{\mathcal L} & = & - M^2 \varepsilon_{s,\infty} \frac{\beta^3 \exp(\beta)}{2 + 2 \beta + \beta^2}  \sum_{\epsilon_\sigma = \pm 1} \int^{\xi_\mathrm{mb}}_{\xi_\mathrm{ph}} d \bar \xi \varepsilon_p^\prime (\bar \xi,\epsilon_\sigma) \exp[- \beta \varepsilon_p(\bar \xi,\epsilon_\sigma)] \left[ \frac{\epsilon_\sigma}{2} \lambda_p^2(\bar \xi,\epsilon_\sigma) + \alpha \varepsilon_p(\bar \xi, \epsilon_\sigma) \lambda_p(\bar \xi,\epsilon_\sigma) \right],
\end{eqnarray}
\end{subequations}
where $\varepsilon_p^\prime (\xi,\epsilon_\sigma) = \partial \varepsilon_p (\xi,\epsilon_p)/\partial \xi$.
\end{widetext}

The above parametric representation fails for $\alpha = \pm 1$, i.e., for the extreme Kerr metric. For $\epsilon_\sigma \alpha = + 1$ (prograde motion), we have $\xi_\mathrm{ph} = \xi_\mathrm{mb} = 1$, and the lower and upper integration limits in the integrals corresponding to $\epsilon_\sigma = \alpha$ coincide. On the other hand, correct expressions given by Eqs.\ (\ref{ratesepsilon}) yield non zero results for those integrals. This behavior is not a fault of Boyer-Lindquist coordinates, as it is discussed in a beautiful article by Jacobson \cite{Jacobson}, but it is a failure of parametrization (\ref{parametric}) for the extreme Kerr metric. Since for $\alpha = \pm 1$ and $\epsilon_\sigma \alpha = +1$ we have $\lambda_c (\varepsilon,\epsilon_\sigma) = \varepsilon$ (cf.~Fig.~\ref{fig:groebner3}), integrals appearing in Eqs.\ (\ref{ratesepsilon}) can be easily evaluated. We get in this case (assuming for simplicity positive signs $\alpha = +1$, $\epsilon_\sigma = +1$)
\begin{subequations}
\begin{eqnarray}
    \label{Mdota1eps1}
    \dot{M}|_{\alpha = +1, \epsilon_\sigma =  + 1} & = & M \rho_{s,\infty}, \\
    \dot{\mathcal{E}}|_{\alpha = + 1, \epsilon_\sigma = + 1} & = & M \varepsilon_{s,\infty}, \\
    \dot{\mathcal{L}}|_{\alpha = + 1, \epsilon_\sigma = + 1} & = & \frac{3}{2} M^2 \varepsilon_{s,\infty}.
\end{eqnarray}
\end{subequations}

Although, $\dot M$ and $\dot{\mathcal{E}}$ appear on a similar footing, the quantity responsible for the growth of the black hole mass $M$ is the energy accretion rate $\dot{\mathcal{E}}$ rather than $\dot M$. This is shown in spherical symmetry and in the context of perfect fluids in \cite{machmalec2022} (see also \cite{machmalec2006}).

\begin{figure*}
    \centering
    \includegraphics[width=\columnwidth]{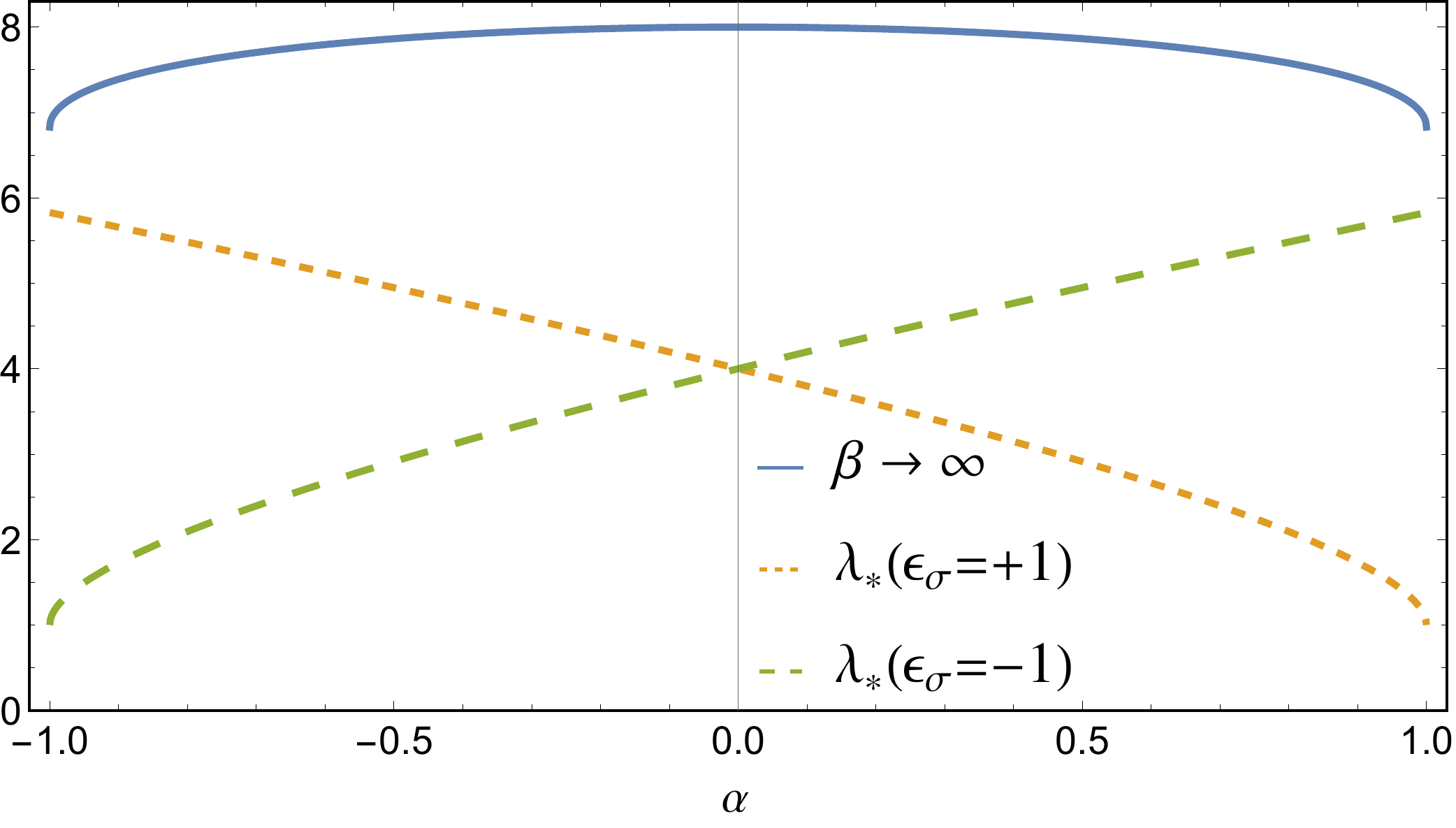}
    \includegraphics[width=\columnwidth]{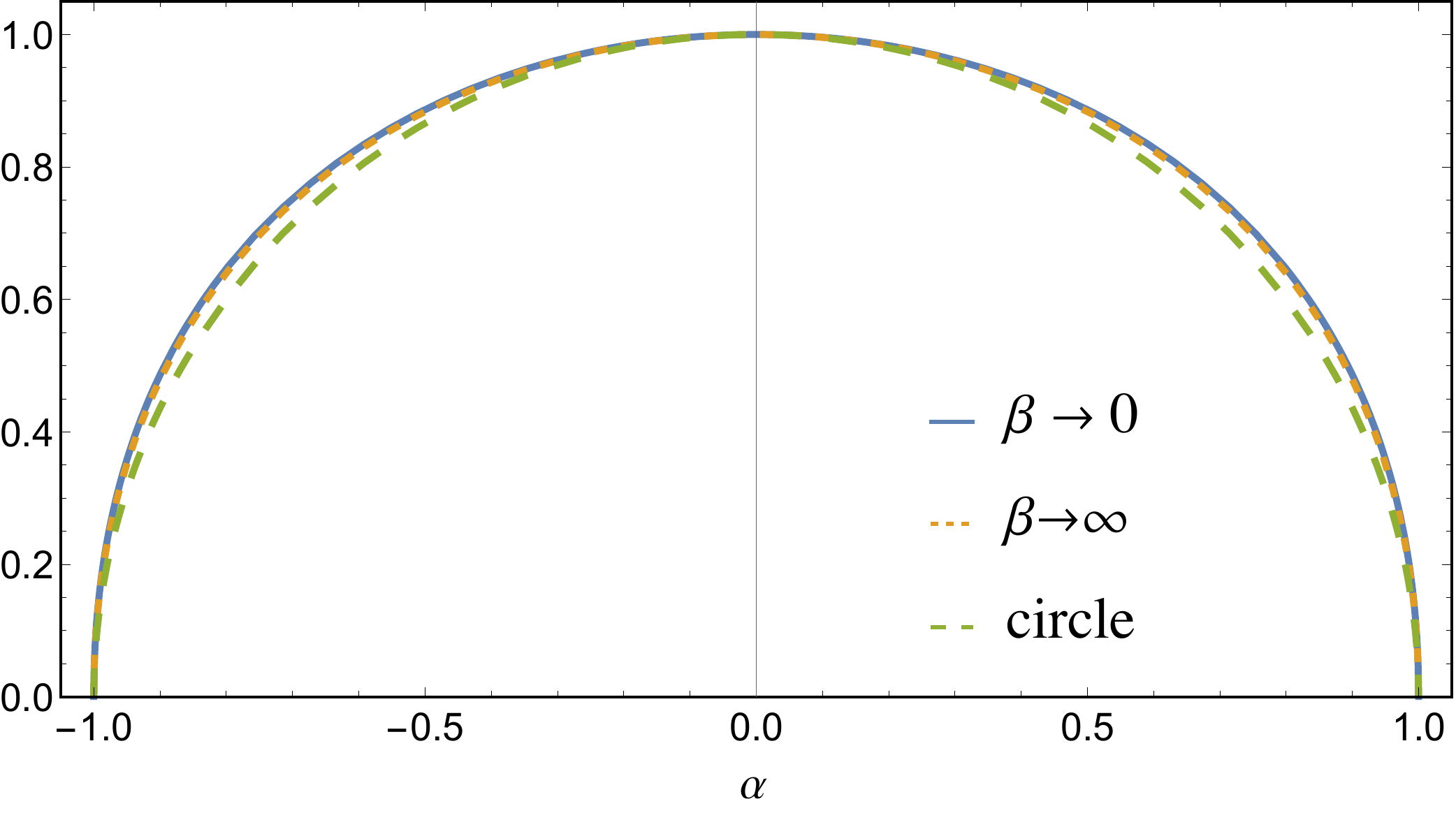}
    \caption{Accretion rate $\dot M$ vs.\ black hole spin parameter $\alpha$. The abscissa in the left panel shows $\lim_{\beta \to \infty} \dot M/(M \rho_{s,\infty}) = \sum_{\epsilon_\sigma = \pm 1} \lambda_\ast(\epsilon_\sigma)$ (solid blue line), $\lambda_\ast(\epsilon_\sigma = +1)$ (dotted orange line), and $\lambda_\ast(\epsilon_\sigma = -1)$ (dashed green line). The right panel depicts accretion rates $\dot M$, rescaled according to Eq.\ (\ref{mdotrescaling}). Blue and orange graphs correspond to the limits $\beta \to 0$ and $\beta \to \infty$, respectively.  A graph of the half circle $\sqrt{1 - \alpha^2}$ (dashed green line) is shown for comparison.}
    \label{fig:Mdot}
\end{figure*}

\begin{figure}
    \centering
    \includegraphics[width=\columnwidth]{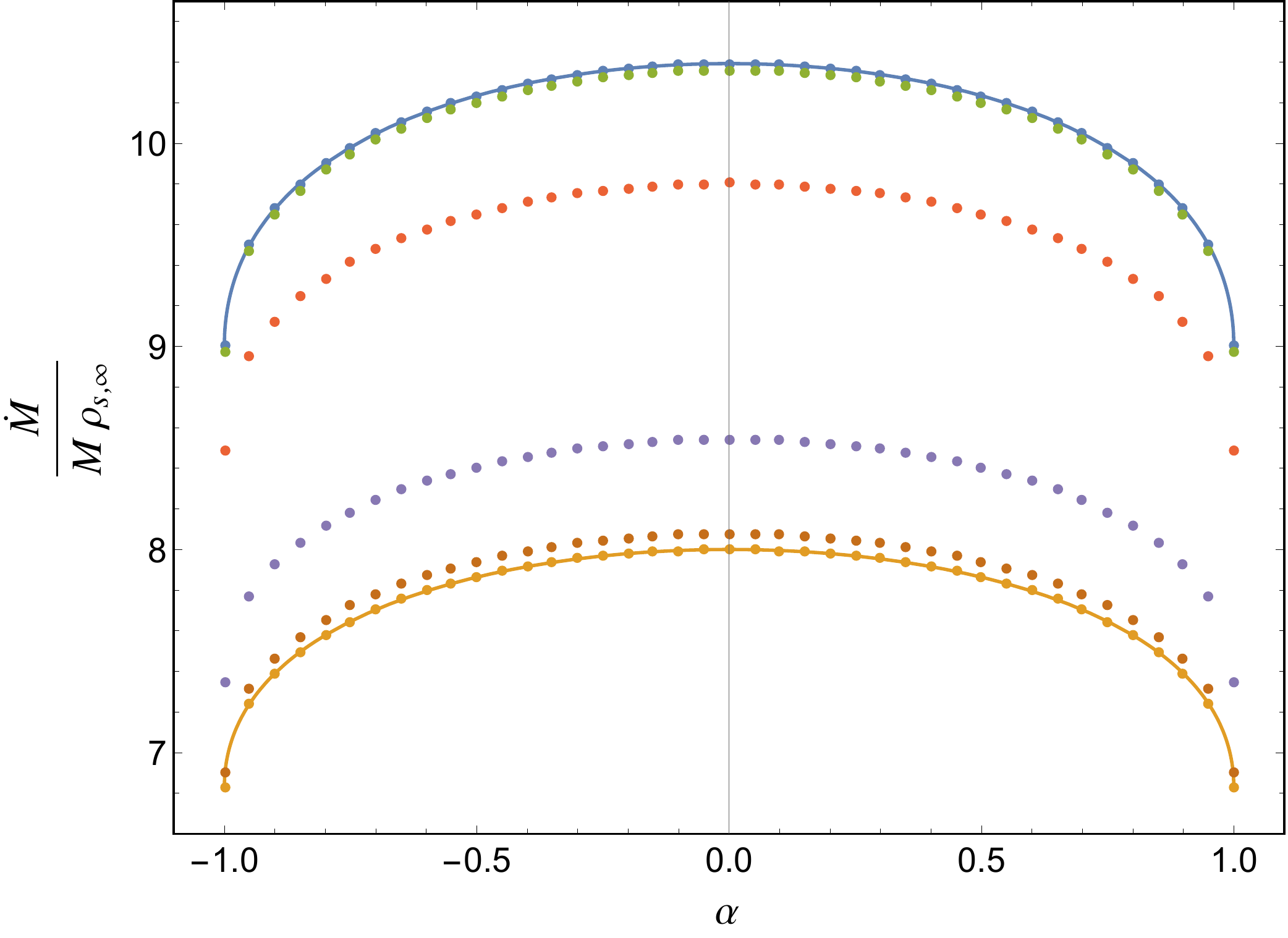}
    \caption{$\dot M/(M \rho_{s,\infty})$ versus $\alpha$. The dots represent numerically computed values. Solid lines depict the limits $\beta \to 0$ (blue line) and $\beta \to \infty$ (orange line), given by Eqs.\ (\ref{Mdot_hot}) and (\ref{Mdot_cold}), respectively. Numerical data are computed for $\beta = 1/4000$, $1/10$, 1, 10, 100, and 10000. }
    \label{fig:mdotalpha}
\end{figure}

\begin{figure}
    \centering
    \includegraphics[width=\columnwidth]{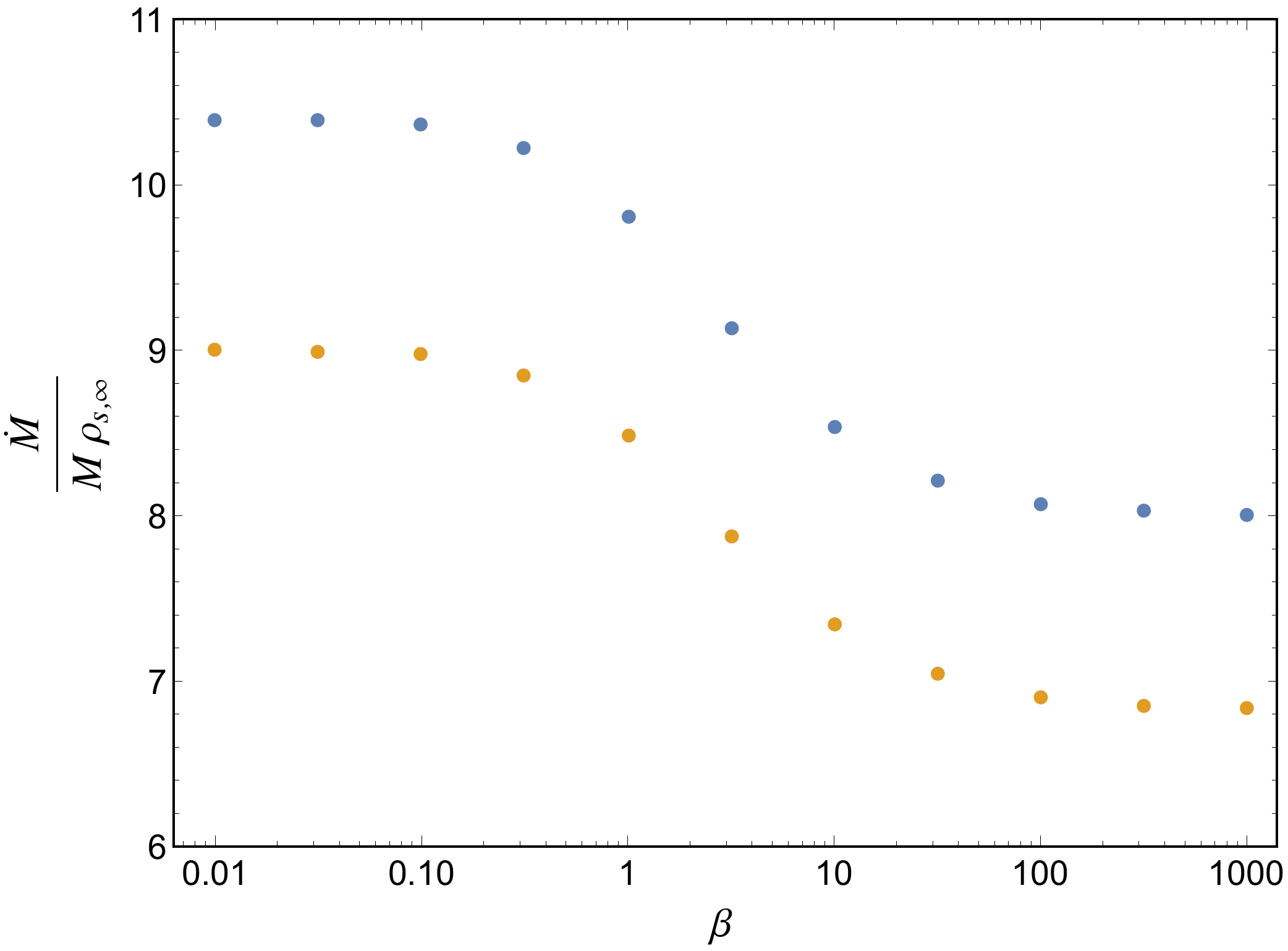}
    \caption{$\dot M/(M \rho_{s,\infty})$ versus $\beta$. Blue dots correspond to $\alpha = 0$. Orange dots correspond to $\alpha = \pm 1$. For $\alpha=0$ the data interpolate between $6 \sqrt{3}$ and $8$. For $\alpha = \pm 1$, between 9 and $2(2+\sqrt{2})$. The inflection point is located around $\beta\simeq 2.8$.}
    \label{fig:mdotbeta}
\end{figure}

\begin{figure}
    \centering
    \includegraphics[width=\columnwidth]{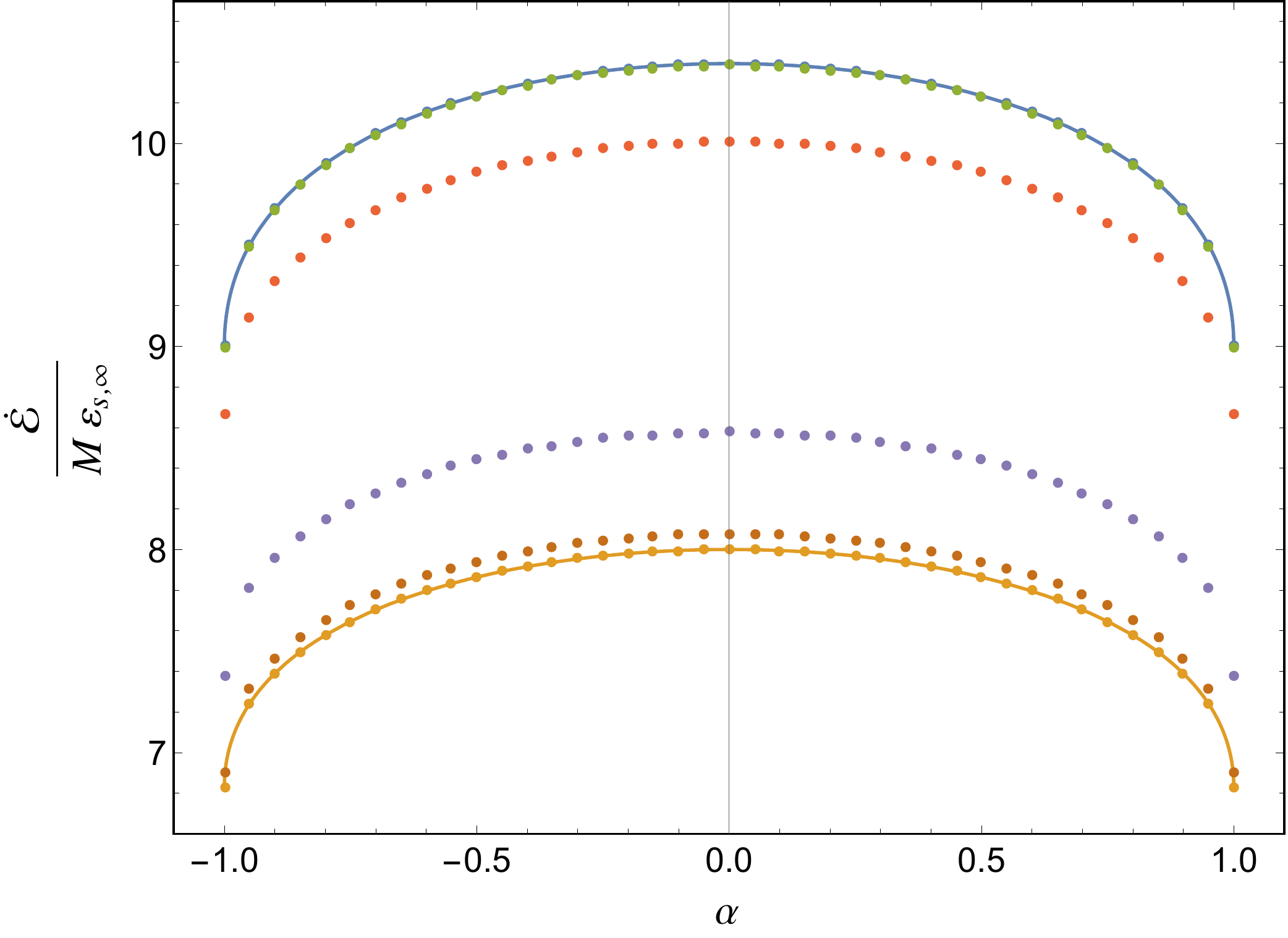}
    \caption{$\dot{\mathcal{E}}/(M \varepsilon_{s,\infty})$ versus $\alpha$. The dots represent numerically computed values. Solid lines depict the limits $\beta \to 0$ (blue line) and $\beta \to \infty$ (orange line), given by Eqs.\ (\ref{Edot_hot}) and (\ref{Edot_cold}), respectively. Numerical data are computed for $\beta = 1/1000$, $1/10$, 1, 10, 100, and 10000.}
    \label{fig:edotalpha}
\end{figure}

\begin{figure}
    \centering
    \includegraphics[width=\columnwidth]{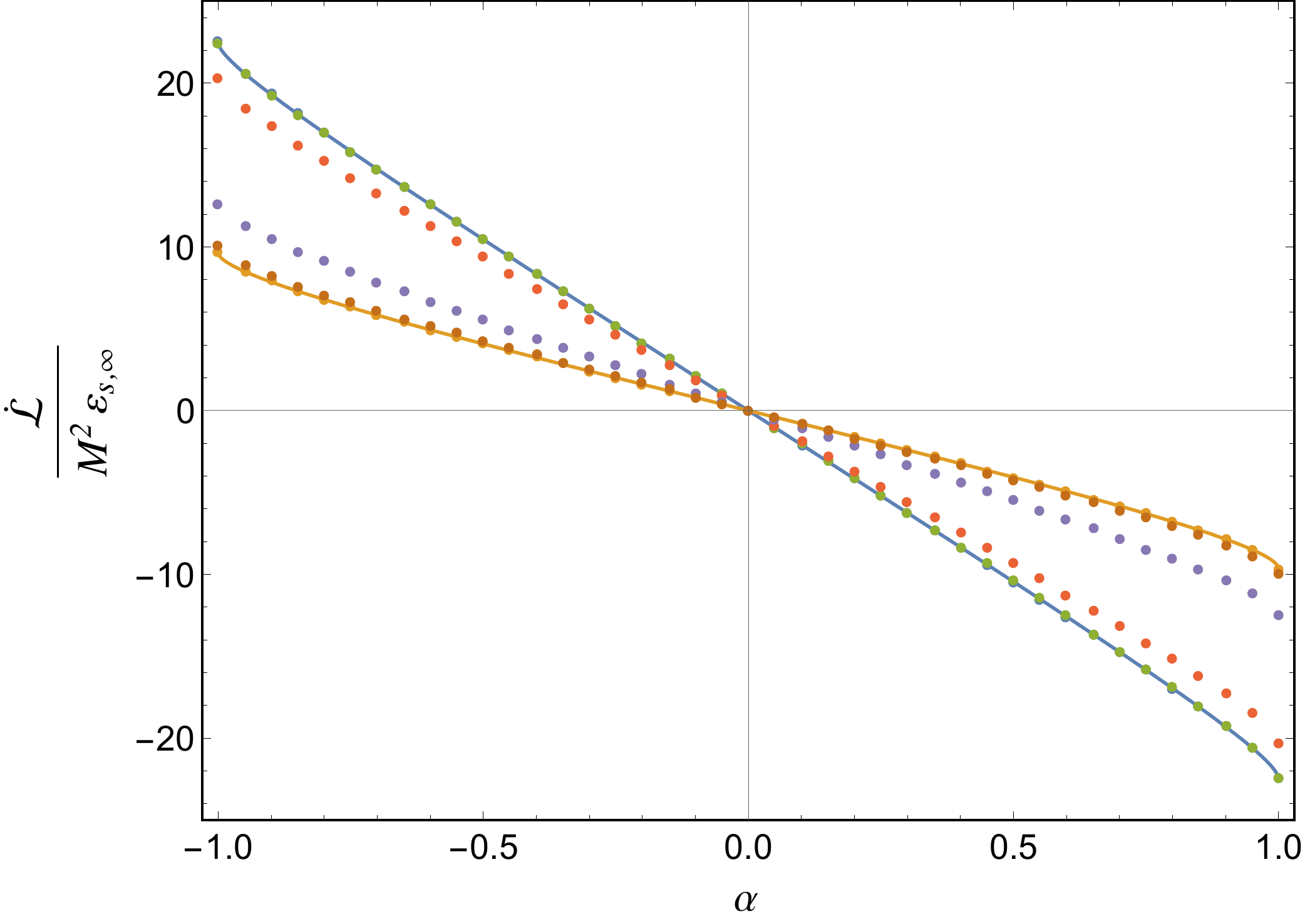}
    \caption{$\dot{\mathcal{L}}/(M^2 \varepsilon_{s,\infty})$ versus $\alpha$. The dots represent numerically computed values. Solid lines depict the limits $\beta \to 0$ (blue line) and $\beta \to \infty$ (orange line), given by Eqs.\ (\ref{Ldot_hot}) and (\ref{Ldot_cold}), respectively. Numerical data are computed for $\beta = 1/100$, $1/10$, 1, 10, 100, and 1000. Note a nearly linear behavior for moderate values of $\alpha$; the deviation
    from a first order series expansion around $\alpha=0$ is of the order of 3.5\% at $\alpha = \pm 0.5$.
    \label{fig:ldotalpha}
    }
\end{figure}

While the integrals in Eqs.~\ref{dotMEL_param}  can be computed numerically, relatively simple analytic expressions can be obtained in the limits of $\beta \to \infty$ (zero asymptotic temperature) and $\beta \to 0$ (ultrarelativistic particles).

The asymptotic behavior of $\dot M$ for $\beta \to \infty$ (zero asymptotic temperature) follows from Watson's lemma on asymptotic expansions of Laplace-type integrals \cite{Watson}. It gives
\begin{equation}
\int_1^\infty e^{-\beta \varepsilon} \lambda_c(\varepsilon, \epsilon_\sigma) \; d \varepsilon
\sim
\frac{e^{-\beta}}{\beta} \sum_{n=0}^{\infty} \frac{1}{\beta^n} \left. \frac{\partial^n \lambda_c(\varepsilon,\epsilon_\sigma)}{\partial \varepsilon^n} \right|_{\varepsilon=1}.
\end{equation}
Note that the lower integration limit is 1 instead of zero, which leads to an additional factor $e^{-\beta}$, as compared to the standard textbook version of Watson's lemma.
Since $\lambda_c(1,\epsilon_\sigma) = \lambda_\ast(\epsilon_\sigma) \equiv 2 - \epsilon_\sigma \alpha + 2 \sqrt{1 - \epsilon_\sigma \alpha}$, and
\begin{equation}
    \left. \frac{\partial \lambda_c(\varepsilon,\epsilon_\sigma)}{\partial \varepsilon} \right|_{\varepsilon=1} = \lambda_\ast(\epsilon_\sigma)^\frac{3}{2},
\end{equation}
we get, up to the first order,
\begin{equation}
   \int_1^\infty e^{-\beta \varepsilon} \lambda_c(\varepsilon, \epsilon_\sigma) \; d \varepsilon
\sim \frac{e^{-\beta}}{\beta} \left[ \lambda_\ast(\epsilon_\sigma) + \frac{1}{\beta} \lambda_\ast (\epsilon_\sigma)^\frac{3}{2} \right]. 
\end{equation}
Since
\begin{equation}
    \sum_{\epsilon_\sigma = \pm 1} \lambda_\ast(\epsilon_\sigma) = 4 + 2 \sqrt{1 - \alpha}
 + 2 \sqrt{1 + \alpha},
\end{equation}
we obtain the asymptotic expansion of $\dot M$ in the form
\begin{eqnarray}
    \dot M & \sim & M \rho_{s,\infty} \frac{\beta}{1 + \beta} \bigg\{ 4 + 2 \sqrt{1 - \alpha}
 + 2 \sqrt{1 + \alpha} \nonumber \\
 &&  + \frac{1}{\beta} \left[ \left( 2 - \alpha + 2 \sqrt{1 - \alpha} \right)^\frac{3}{2} \right. \nonumber \\
 &&  \left. + \left( 2 + \alpha + 2 \sqrt{1 + \alpha} \right)^\frac{3}{2} \right]   \bigg\}.
\end{eqnarray}
The limit for $\beta \to \infty$ reads
\begin{equation}
\label{Mdot_cold}
\lim_{\beta \to \infty} \dot{M} =  2 M \rho_{s,\infty} \left( 2+\sqrt{1-\alpha} + \sqrt{1+\alpha} \right).
\end{equation}
The above expression can also be written in the following trigonometric form
\begin{equation}
\lim_{\beta \to \infty} \dot{M} = 8 M \rho_{s,\infty} \cos^2 \left(\frac{\arcsin{\alpha}}{4}\right).
\end{equation}

In the case of $\dot{\mathcal E}$, we get, for $\beta \to \infty$, a similar expansion
\begin{eqnarray}
   \lefteqn{ \int_1^\infty \exp(- \beta \varepsilon) \varepsilon \lambda_c(\varepsilon,\epsilon_\sigma) d \varepsilon} \nonumber \\
   && \sim \frac{e^{-\beta}}{\beta} \sum_{n=0}^\infty \frac{1}{\beta^n} \left. \frac{\partial^n}{\partial s^n} [(s+1) \lambda_c(s+1,\epsilon_\sigma)] \right|_{s=0}.
\end{eqnarray}
This yields the limit
\begin{eqnarray}
\label{Edot_cold}
 \lim_{\beta \to \infty} \dot{\mathcal{E}} & = & 2 M \rho_{s,\infty} \left(2 + \sqrt{1 - \alpha} + \sqrt{1 + \alpha} \right) \nonumber \\
    & = & 2 M \varepsilon_{s,\infty} \left( 2 + \sqrt{1 - \alpha} + \sqrt{1 + \alpha} \right).
\end{eqnarray}
Clearly, the expressions for $\dot M$ and $\dot {\mathcal{E}}$ coincide for $\beta \to \infty$, as expected.

A similar calculation for $\dot{\mathcal{L}}$ leads to an asymptotic expansion for large values of $\beta$ of the form
\begin{eqnarray}
   \lefteqn{ \sum_{\epsilon_\sigma = \pm 1} \int_1^\infty d \varepsilon \exp(- \beta \varepsilon) \left[ \frac{\epsilon_\sigma}{2} \lambda_c^2(\varepsilon,\epsilon_\sigma) + \alpha \varepsilon \lambda_c(\varepsilon,\epsilon_\sigma) \right]} \nonumber \\
   && \sim \frac{4 e^{-\beta}}{\beta} \left( \sqrt{1 - \alpha} - \alpha - \sqrt{1 + \alpha} \right).
\end{eqnarray}
Thus,
\begin{equation}
   \lim_{\beta \to \infty} \dot{\mathcal{L}} = 4 M^2 \rho_{s,\infty} \left( \sqrt{1 - \alpha} - \alpha - \sqrt{1 + \alpha} \right)
\label{Ldot_cold}
\end{equation}
or, using trigonometric functions,
\begin{equation}
\lim_{\beta \to \infty} \dot{\mathcal{L}} =
-16 M^2 \rho_{s,\infty} \cos^2 \left(\frac{\arcsin{\alpha}}{4} \right)
\sin \left( \frac{\arcsin{\alpha}}{2} \right).
\end{equation}

To find the limits for $\beta \to 0$, we first observe that the integrand of, e.g., \eqref{dotM_param} has a sharp maximum near the beginning of the integration interval, i.e., near $\xi_\mathrm{ph}$. Our method is based on identifying the leading terms in the integrand function and estimating the remainder. The latter can be done using the general Lebesgue dominated convergence theorem \cite{Royden} (page 89, Theorem\ 19). It is a version of the standard Lebesgue dominated convergence theorem in which the dominating function is replaced by a suitable series of functions. We will illustrate this method on 2 particular examples, before proceeding with the general result.

Let us start with the $\alpha = 0$ case. The goal is to estimate the integral
\begin{eqnarray}
I_1 & = & -\frac{e^{\beta } \beta ^2 }{1+\beta } \int_{\xi_\mathrm{ph}}^{\xi_\mathrm{mb}} d \xi  \varepsilon_p^\prime (\xi) \exp [ - \beta \varepsilon_p(\xi) ] \lambda_p(\xi) \\
& = & -\frac{e^{\beta } \beta ^2 }{1+\beta }
 \int _3^4\frac{\xi - 6}{2 (\xi -3)^2 \sqrt{\xi }} \exp \left[ -\frac{\beta  (\xi -2)}{\sqrt{(\xi -3) \xi }}  \right] d\xi. \nonumber
\end{eqnarray}
It cannot be evaluated analytically. However, since the maximum of the integrand is, in the limit, very close to $\xi = 3$, one can try to substitute $\xi = 3$ everywhere except for terms $\xi - 3$, i.e., divergent ones. To be more precise, we split the integral $I_1$ into two parts $I_1 = I_2 + I_3$, where
\begin{equation}
I_2 = \frac{e^{\beta } \beta ^2}{\beta +1}  \int _3^4 \frac{\sqrt{3}}{2 (\xi -3)^2} \exp \left( -\frac{\beta}{\sqrt{3} \sqrt{\xi -3}}  \right) d\xi
\label{I2}
\end{equation}
and
\begin{equation}
    I_3 = - \frac{e^{\beta } \beta ^2 }{1+\beta } \int_3^4 G_1(\xi,\beta) d\xi,
\end{equation}
with
\begin{eqnarray}
    G_1(\xi,\beta) & = & \frac{\sqrt{3}}{2 (\xi -3)^2} \exp \left( -\frac{\beta }{\sqrt{3} \sqrt{\xi -3}} \right) \\
    && -  \frac{6 - \xi}{2 (\xi -3)^2 \sqrt{\xi }} \exp \left[ -\frac{\beta  (\xi -2)}{\sqrt{(\xi -3) \xi }} \right]. \nonumber
\end{eqnarray}
The integral $I_2$ can be evaluated. We have
\begin{equation}
    I_2 = \frac{\sqrt{3} \left(\sqrt{3} \beta +3\right)}{\beta +1} \exp \left( \beta -\frac{\beta }{\sqrt{3}} \right).
\label{I2computed}
\end{equation}
Moreover,
\begin{equation}
    \lim_{\beta \to 0} I_2 = 3 \sqrt{3}.
\end{equation}
We now show that $\lim_{\beta \to 0} I_3 = 0$. Noticing elementary inequalities
\begin{equation}
\sqrt{3} \ge \frac{6 - \xi}{\sqrt{\xi}}
\end{equation}
and
\begin{equation}
\frac{\xi - 2}{\sqrt{\xi}} \ge \frac{1}{\sqrt{3}},
\end{equation}
valid for $3 \le \xi \le 4$, we see that the integrand $G_1(\xi,\beta)$ is non negative. It is also easy to check that it tends to zero, as $\xi \to 3$ from above. Clearly,
\begin{equation}
\label{ineqG}
    G_1(\xi,\beta) < \frac{\sqrt{3}}{2 (\xi -3)^2} \exp \left( -\frac{\beta }{\sqrt{3} \sqrt{\xi -3}} \right)
\end{equation}
in the range $3 < \xi < 4$. The right-hand side of inequality (\ref{ineqG}) is integrable, as follows from Eqs.\ (\ref{I2}) and (\ref{I2computed}). We may now apply the general Lebesgue dominated convergence theorem \cite{Royden} to the integral $I_3$. Since
\begin{equation}
    \lim_{\beta \to 0} \frac{e^{\beta } \beta ^2 }{1+\beta } G_1(\xi,\beta) = 0,
\end{equation}
we also have $\lim_{\beta \to 0} I_3 = 0$. Thus $\lim_{\beta \to 0} I_1 = \lim_{\beta \to 0} I_2 = 3 \sqrt{3}$, and finally
\begin{equation}
\lim_{\beta \to 0} \left( \dot{M} \Big|_{\alpha=0} \right) = 6 \sqrt{3} M \rho_{s,\infty}.
\end{equation}
The additional overall factor 2 in the result is due to the fact that for $\alpha = 0$ the integrals corresponding to $\epsilon_\sigma = \pm 1$ coincide.

As the second example we take $\alpha=1, \epsilon_\sigma = -1$, i.e., the extreme Kerr metric and retrograde orbits. The integral to estimate is now
\begin{eqnarray}
    I_4 & = & \int_{\xi_\mathrm{ph}}^{\xi_\mathrm{mb}} d  \xi \varepsilon_p^\prime ( \xi,-1) \exp [ - \beta \varepsilon_p( \xi,-1) ] \lambda_p( \xi,-1) \nonumber \\
    & = & \int_4^{3 + 2 \sqrt{2}} d \xi \frac{\sqrt{\xi} - 3}{2 \xi (\sqrt{\xi} - 2)^2} e^{ - \frac{\beta (\xi - \sqrt{\xi} - 1)}{\xi^{3/4} \sqrt{\sqrt{\xi} - 2}}},
\end{eqnarray}
where we deliberately skip the term $- e^\beta \beta^2/(1 + \beta)$, to shorten the notation. This integral can be expressed as a sum $I_4 = I_5 + I_6$, where
\begin{equation}
\label{I1secondcase}
    I_5 = - \int_4^{3 + 2 \sqrt{2}} \frac{d \xi}{8 (\sqrt{\xi} - 2)^2} \exp \left( - \frac{\beta}{2 \sqrt{2} \sqrt{\sqrt{\xi} - 2}} \right)
\end{equation}
and
\begin{equation}
    I_6 = \int_4^{3 + 2 \sqrt{2}} G_2(\xi,\beta) d \xi,
\end{equation}
with
\begin{eqnarray}
    G_2(\xi,\beta) & = & \frac{1}{8 (\sqrt{\xi} - 2)^2} \exp \left(- \frac{\beta}{2 \sqrt{2} \sqrt{\sqrt{\xi} - 2}} \right) \\
    && - \frac{3 - \sqrt{\xi}}{2 \xi (\sqrt{\xi} - 2)^2} \exp \left[ - \frac{\beta(\xi - \sqrt{\xi} - 1)}{\xi^{3/4} \sqrt{\sqrt{\xi} - 2}} \right]. \nonumber
\end{eqnarray}
We have
\begin{equation}
\label{G2estimate}
    0 \le G_2(\xi,\beta) \le \frac{1}{8 (\sqrt{\xi} - 2)^2} \exp \left(- \frac{\beta}{2 \sqrt{2} \sqrt{\sqrt{\xi} - 2}} \right).
\end{equation}
The first inequality follows again from two elementary inequalities
\begin{equation}
    \frac{1}{4} \ge \frac{3 - \sqrt{\xi}}{\xi}
\end{equation}
and
\begin{equation}
    \frac{1}{2 \sqrt{2}} \le \frac{\xi - \sqrt{\xi} - 1}{\xi^{3/4}},
\end{equation}
valid in the range $4 \le \xi \le 3 + 2 \sqrt{2}$. The second inequality in Eq.\ (\ref{G2estimate}) is obvious, as $3 - \sqrt{\xi} > 0$ for $4 \le \xi \le 3 + 2 \sqrt{2}$. The integral $I_5$ can be evaluated with the result
\begin{eqnarray}
I_5 & = & - \frac{2 \sqrt{2 \left(1+\sqrt{2}\right)} \beta +8}{\beta ^2} \exp \left( -\frac{\beta}{2} \sqrt{\frac{1}{2}+\frac{1}{\sqrt{2}}} \right) \nonumber \\
&& + \frac{1}{2}
   \text{Ei}\left(-\frac{\beta}{2} \sqrt{\frac{1}{2}+\frac{1}{\sqrt{2}}} \right).
\end{eqnarray}
Thus, $G_2(\xi,\beta)$ is bounded by an integrable function. We have
\begin{equation}
    \lim_{\beta \to 0} \left[ \frac{\beta^2 e^\beta}{1 + \beta} I_5 \right] = -8
\end{equation}
and
\begin{equation}
    \lim_{\beta \to 0} \left[ \frac{\beta^2 e^\beta}{1 + \beta} G_2(\xi,\beta) \right] = 0.
\end{equation}
By the generalized Lebesgue theorem
\begin{equation}
    \lim_{\beta \to 0} \left[ \frac{ \beta^2 e^\beta}{1+\beta} I_6 \right] = 0,
\end{equation}
and consequently
\begin{equation}
    \lim_{\beta \to 0} \left[ \frac{\beta^2 e^\beta}{1 + \beta} I_4 \right] = -8.
\end{equation}
Thus,
\begin{equation}
\lim_{\beta \to 0} \left( \dot{M} \Big|_{\alpha=1, \epsilon_\sigma=-1} \right) = 8 M \rho_{s,\infty}.
\end{equation}
The corresponding result for $\alpha = 1$ and $\epsilon_\sigma = +1$ is given by Eq.\ (\ref{Mdota1eps1}), which is independent of $\beta$.

Repeating the above procedure for any $-1 < \alpha < 1$ gives, with the help of a
computer algebra system (we use Wolfram Mathematica \cite{Mma}), the following formula
\begin{widetext}
\begin{eqnarray}
\lim_{\beta \to 0} \dot{M} & = & M \rho_{s,\infty}
\Biggl[
\left(2+\sqrt[3]{-\alpha +i \sqrt{1-\alpha ^2}} \left(-i \sqrt{1-\alpha ^2}+\sqrt[3]{-\alpha +i \sqrt{1-\alpha ^2}}-\alpha
   \right)\right)^{3/2} \nonumber \\
&&   +\left(2+\sqrt[3]{\alpha +i \sqrt{1-\alpha ^2}} \left(-i \sqrt{1-\alpha ^2}+\sqrt[3]{\alpha +i \sqrt{1-\alpha ^2}}+\alpha
   \right)\right)^{3/2}
   \Biggr].
\end{eqnarray}
\end{widetext}
An alternative, explicitly real form of this expression can be written as
\begin{equation}
\lim_{\beta \to 0} \dot{M}  =  6 \sqrt{3} M \rho_{s,\infty} \cos \left( \frac{1}{3} \arcsin{\alpha} \right).
\label{Mdot_hot}
\end{equation}

In a similar way, one can compute the limit of $\dot{\mathcal{E}}$ for $\beta \to 0$. The result reads
\begin{equation}
\label{Edot_hot}
\lim_{\beta \to 0}\dot{\mathcal{E}} = 6 \sqrt{3} M \varepsilon_{s,\infty} \cos \left(\frac{1}{3} \arcsin{\alpha}\right).
\end{equation}
This is quite surprising. In addition to the equality
\begin{equation}
    \lim_{\beta \to \infty} \frac{\dot M}{M \rho_{s,\infty}} = \lim_{\beta \to \infty} \frac{\dot{\mathcal{E}}}{M \varepsilon_{s,\infty}},
\end{equation}
established so far, we now also have
\begin{equation}
    \lim_{\beta \to 0} \frac{\dot M}{M \rho_{s,\infty}} = \lim_{\beta \to 0} \frac{\dot{\mathcal{E}}}{M \varepsilon_{s,\infty}}.
\end{equation}

Finally, for $\dot{\mathcal{L}}$ we get
\begin{eqnarray}
\lim_{\beta \to 0}\dot{\mathcal{L}} & = &  -6 \sqrt{3} M^2 \varepsilon_{s,\infty}
 \sin \left(\frac{2 \arcsin{\alpha} }{3}\right) \nonumber \\
 && \times \left[ 2+\cos \left(\frac{2 \arcsin{\alpha} }{3}\right)\right].
 \label{Ldot_hot}
\end{eqnarray}
In a good approximation, the dependence of $\dot{\mathcal{L}}/(M^2 \varepsilon_{s,\infty})$ on $\alpha$ is just linear (cf. Fig.~\ref{fig:ldotalpha}). The coefficient
\begin{equation}
    \delta \equiv \frac{d}{d\alpha} \left. \left( \frac{\dot{\mathcal{L}}}{M^2 \varepsilon_{s,\infty}} \right) \right|_{\alpha = 0}
\end{equation}
varies from $\delta = -8$ in the limit $\beta \to \infty$ to $\delta = -12 \sqrt{3}$ for $\beta \to 0$.

Figures \ref{fig:mdotalpha}--\ref{fig:ldotalpha} show numerical values of the accretion rates $\dot M$, $\dot{\mathcal{E}}$, and $\dot{\mathcal{L}}$ together with the corresponding limits, obtained for $\beta \to \infty$ and $\beta \to 0$. Figure \ref{fig:mdotalpha} depicts the dependence of the rest-mass accretion rate on the black hole spin parameter $\alpha$. Different colors represent solutions corresponding to different values of the parameter $\beta$. The first observation is that $\dot M/(M \rho_{s,\infty})$ decreases with $|\alpha|$. The dependence of $\dot M/(M \rho_{s,\infty})$ on $\alpha$ is, in fact, similar for different values of $\beta$ and can be illustrated by the limiting cases $\beta \to 0$ and $\beta \to \infty$, given by Eqs.\ (\ref{Mdot_hot}) and (\ref{Mdot_cold}). After a suitable rescaling, the graphs of $\dot M/(M \rho_{s,\infty})$ vs.\ $\alpha$ have an almost circular shape, as shown in Fig.\ \ref{fig:Mdot}. The quantity plotted in this figure (the rescaled $\dot M (\beta,\alpha)/(M \rho_{s,\infty})$) is
\begin{equation}
\label{mdotrescaling}
\frac{\dot{M}(\beta, \alpha) - \dot{M}(\beta, 1)}{\dot{M}(\beta, 0) - \dot{M}(\beta, 1)}
\end{equation}
for fixed values of $M$ and $\rho_{s,\infty}$. Upon such a rescaling, the difference between the values corresponding to $\beta \to 0$ and $\beta \to \infty$ is less than 1\%, and it is barely visible on the plot.  The nearly circular shape can be explained by the zero temperature limit, $\beta \to \infty$, in which $\dot M/(M \rho_{s,\infty})$ is just the sum $\lambda_\ast(\epsilon_\sigma = +1) + \lambda_\ast(\epsilon_\sigma = -1) = 4 + 2 \sqrt{1 - \alpha} + 2 \sqrt{1 + \alpha}$.

The obtained dependence of $\dot M/(M \rho_{s,\infty})$ on $\alpha$ confirms our previous results reported in \cite{cieslik}, where two of us investigated a spherical model of the steady accretion of the collisionless Vlasov gas on the Reissner-Nordstr\"{o}m black hole. In this work the charged black hole was treated as a toy model of a rotating one, allowing for a simple analysis in spherical symmetry. We have found that the rest-mass accretion rate decreases with the charge parameter (see Fig.\ 7 in \cite{cieslik}). 

Another fact is related to the dependence of $\dot M/(M \rho_{s,\infty})$ on $\beta$. We see both from Figs.\ \ref{fig:mdotalpha} and \ref{fig:mdotbeta} that the ratio $\dot M/(M \rho_{s,\infty})$ is actually decreasing with $\beta$, i.e., it increases with the asymptotic temperature of the gas. This is surprising, as the situation known for the spherically symmetric accretion of the Vlasov gas onto the Schwarzschild black hole \cite{Olivier1,Olivier2} or the Reissner-Nordstr\"{o}m black hole \cite{cieslik} is exactly opposite. In the latter case, the mass accretion rate $\dot M$, normalized by $M^2$ and the asymptotic rest-mass density, decreases with the asymptotic temperature of the gas. Moreover, it is divergent for $\beta \to \infty$. It turns out that the factor responsible for the reversed behavior in our case is the parametrization of $\dot M$ by the asymptotic surface rest-mass density, and its dependence on the parameter $\beta$. We will return to the discussion of this behavior in the subsequent section.

The energy accretion rate $\dot{\mathcal E}$ exhibits a very simlar dependence on $\alpha$ and $\beta$, provided that it is normalized by the asymptotic surface energy density $\varepsilon_{s,\infty}$, and not by $\rho_{s,\infty}$. This behavior is illustrated in Fig.\ \ref{fig:edotalpha}.

The most interesting outcome of our study concerns, in our view, the angular momentum accretion rate. The dependence of $\dot{\mathcal L}/(M^2 \varepsilon_{s,\infty})$ is shown in Fig.\ \ref{fig:ldotalpha}. The sign of $\dot{\mathcal L}$ is always opposite to the sign of the black hole spin parameter $\alpha$, meaning that the accretion process slows down the black hole rotation. This seems to be reasonable---in our model the black hole is placed in a large (infinite) reservoir of the gas with essentially no net angular momentum. A steady accretion of the gas can, therefore, decrease the black hole angular momentum. Moreover, the dependence of $\dot{\mathcal L}/(M^2 \varepsilon_{s,\infty})$ on $\alpha$ is to a good approximation linear, except for very high values of $|\alpha|$. The dependence of $\dot{\mathcal L}/(M^2 \varepsilon_{s,\infty})$ on the parameter $\beta$ is qualitatively similar to the dependence of the rest-mass and energy accretion rates on $\beta$. The values interpolate between two limits obtained for $\beta \to 0$ and $\beta \to \infty$, given by Eqs.\ (\ref{Ldot_hot}) and (\ref{Ldot_cold}).

\section{Discussion}
\label{sec:discussion}

We have investigated a model of a geometrically thin, stationary accretion disk composed of the collisionless Vlasov gas moving in the equatorial plane of the Kerr spacetime. This model serves as a two dimensional kinetic analogue of the Bondi (or Michel) type accretion---at infinity the gas is assumed to be homogeneous and at rest, while the motion in the vicinity of the black hole is induced by the gravitational attraction.

One of subtle points in our analysis is related to the parametrization of solutions. The normalization constant $A$ appearing in the distribution function (\ref{f3d}) is not an observable quantity. To obtain a consistent formalism, we express this constant by the asymptotic rest-mass surface density $\rho_{s,\infty}$ or the asymptotic energy surface density $\varepsilon_{s,\infty}$. The relation of these quantities with the asymptotic temperature $T$ or the parameter $\beta$, characteristic for the two dimensional Maxwell-J\"{u}ttner distribution, leads to an unexpected dependence of the rest-mass and energy accretion rates on $\beta$---both $\dot M/(M \rho_{s,\infty})$ and $\dot{\mathcal E}/(M \varepsilon_{s,\infty})$ decrease with increasing $\beta$. The situation known for spherically symmetric models of steady accretion of the collisionless Vlasov gas on Schwarzschild or Reissner-Nordstr\"{o}m black holes is the opposite---the mass accretion rate $\dot M$, normalized by $M^2$ and the asymptotic rest-mass density $\rho_\infty$, grows with $\beta$ \cite{Olivier1,Olivier2,cieslik}. We have checked that replacing the constant $A$ in our formula for $\dot M$ with a relation characteristic for a three dimensional Maxwell-J\"{u}ttner distribution, i.e.,
\begin{equation}
    A = \frac{\beta \rho_\infty}{4 \pi m_0^5 K_2(\beta)},
\end{equation}
[cf.\ Eq.\ (\ref{3dn})], would restore (qualitatively) the same standard behavior. This implies a natural question about the physically relevant parametrization. Note that even a tiny momentum component $p_z$, normal to the equatorial plane, would cause a deviation from the equatorial plane sufficiently far from the black hole and imply a three dimensional asymptotic distribution.

Another aspect in which the dimensionality of our model is clearly visible is related to the changes of the black hole spin implied by the angular momentum accretion rate. Let us estimate these changes assuming a quasistationary approximation. Assume that $\dot{\mathcal{L}}$ is a time derivative of the black hole angular momentum $J = M^2 \alpha$ and that the energy accretion rate $\dot{\mathcal E}$ gives the time derivative of the black hole mass $M$. Thus
\begin{equation}
    \frac{d \alpha}{dt} = \frac{d}{dt} \frac{J}{M^2} = \frac{J}{M^2} \left( \frac{\dot {\mathcal{L}}}{J} - 2 \frac{\dot{\mathcal{E}}}{M} \right) = \alpha \left( \frac{\dot{\mathcal{L}}}{J} - 2 \frac{\dot{\mathcal{E}}}{M} \right).
\end{equation}
In the limit of $\beta \to \infty$ (cold matter) we get, for small values of $\alpha$,
\begin{equation}
    \frac{d\alpha}{dt} \simeq - 24 \rho_{s,\infty} \alpha.
\end{equation}
Therefore, in this limit the rotation of the black hole slows down on the time scale $\tau = c/(24 G \rho_{s,\infty})$ (in SI units). Note that in our two dimensional model the above result does not depend on the black hole mass $M$. As an illustration, we could estimate the time scale $\tau$ for the dark matter component present in the Milky Way. Projecting the local cold dark matter density $\rho_\text{DM} \simeq 5 \times 10^{-22}$~kg/m$^3$ \cite{localDMdensity} on the Galactic plane, we get $\rho_{s,\infty} = 2 r_0 \rho_\text{DM}$, where $r_0 = 50$~kpc is the dark matter halo scale \cite{DMhaloradius}. This gives the time scale $\tau$ of the order of $4 \times 10^9$ years. We emphasize that the above result clearly depends on the dimensionality of the model. Improving of the above estimate would require a fully three dimensional accretion model.

\begin{acknowledgments}
The authors would like to thank Olivier Sarbach for discussions. P.\ M.\ was partially supported by the Polish National Science Centre Grant No.\ 2017/26/A/ST2/00530. A.\ C.\ acknowledges a support of the Faculty of Physics, Astronomy and Applied Computer Science of the Jagiellonian University, grant No.\ N17/MNS/000051.
\end{acknowledgments}

\appendix

\section{4 dimensional calculation}
\label{appendixA}

In this appendix we recover main formulas of this paper, performing a fully 4-dimensional calculation in the phase-space and imposing a restriction to the equatorial plane afterwards.

In the four-dimensional setting, the distribution function has the form $\mathcal F = \mathcal F(t,r,\theta,\varphi,p_t,p_r,p_\theta,p_\varphi)$. The particle current density and the energy momentum tensor are defined as
\begin{equation}
\label{Jmufull}
    \mathcal J_\mu (x) = \int \mathcal F(x,p) p_\mu \mathrm{dvol}_x(p),
\end{equation}
and
\begin{equation}
\label{Tmunufull}
    \mathcal T_{\mu \nu} (x) = \int \mathcal F(x,p) p_\mu p_\nu \mathrm{dvol}_x(p),
\end{equation}
where
\begin{equation}
    \mathrm{dvol}_x(p) = \sqrt{- \mathrm{det} (g^{\mu \nu})} dp_t dp_r dp_\theta dp_\varphi.
\end{equation}
For the motion restricted to the equatorial plane we take
\begin{equation}
\label{twodf}
    \mathcal F(x,p) = \delta(z) F(x,p),
\end{equation}
where $z$ is a coordinate such that the vector $\partial_z$ is normal to the equatorial plane at $z = 0$. In terms of spherical coordinates used in this paper we can assume the standard formula $z = r \cos \theta$. This gives $\delta(z) = \delta(\theta - \pi/2)/r$. Thus,
\begin{subequations}
\label{surfacequant}
\begin{eqnarray}
     \mathcal J_\mu (x) & = & \delta(z) J_\mu(x), \\
     \mathcal T_{\mu \nu} (x) & = &  \delta(z) T_{\mu \nu}(x),
\end{eqnarray}
\end{subequations}
where
\begin{equation}
\label{Jmu4d}
    J_\mu (x) = \int F(x,p) p_\mu \mathrm{dvol}_x(p),
\end{equation}
and
\begin{equation}
\label{Tmunu4d}
    T_{\mu \nu} (x) = \int F(x,p) p_\mu p_\nu \mathrm{dvol}_x(p).
\end{equation}

The distribution function $\mathcal F$ satisfies the Vlasov equation
\begin{eqnarray}
\label{vlasoveq4d}
    \frac{d \mathcal F}{d\tau} & = & \frac{\partial \mathcal F}{\partial x^\mu} \frac{d x^\mu}{d \tau} + \frac{\partial \mathcal F}{\partial p_\nu} \frac{d p_\nu}{d \tau} = \frac{\partial \mathcal F}{\partial x^\mu} \frac{\partial H}{\partial p_\mu} - \frac{\partial \mathcal F}{\partial p_\nu} \frac{\partial H}{\partial x^\nu} \nonumber \\
    & = & \{ H, \mathcal F \} = 0.
\end{eqnarray}

To compute momentum integrals in Eqs.\ (\ref{Jmu4d}) and (\ref{Tmunu4d}), we continue in a way similar to our 3 dimensional calculation and consider a transformation of momentum variables $(p_t,p_r,p_\theta,p_\varphi) \mapsto (m^2, E, l^2, l_z)$, given by
\begin{widetext}
\begin{subequations}
\label{pmutoconst}
\begin{eqnarray}
\label{m2}
m^2 & = & - \left[ g^{tt} (p_t)^2 + 2 g^{t\varphi} p_t p_\varphi + g^{r r} (p_r)^2 + g^{\theta \theta} (p_\theta)^2 + g^{\varphi \varphi} (p_\varphi)^2 \right], \\
E & = & - p_t, \\
\label{l2}
l^2 & = & p_\theta^2 + m^2 a^2 \cos^2 \theta + \left( \frac{p_\varphi}{\sin \theta} + a \sin \theta p_t \right)^2, \\
l_z & = & p_\varphi,
\end{eqnarray}
\end{subequations}
\end{widetext}
where $m^2$ in Eq.\ (\ref{l2}) has to be replaced by the right-hand side of Eq.\ (\ref{m2}). By a straightforward computation, we get the Jacobian determinant
\begin{equation}
    \frac{\partial (m^2,E,l^2,l_z)}{\partial (p_t,p_r,p_\theta,p_\varphi)} = - \frac{4 \Delta p_r p_\theta}{\rho^2} = \pm \frac{4 \sqrt{R} \sqrt{\Theta}}{\rho^2}.
\end{equation}
Consequently
\begin{eqnarray}
    d m^2 dE dl^2 dl_z & = & 4 (m d m)dE (l dl) dl_z  \nonumber \\
    & = & \frac{4 \sqrt{R} \sqrt{\Theta}}{\rho^2} dp_t dp_r dp_\theta dp_\varphi.
\end{eqnarray}
For the motion confined to the equatorial plane $p_\theta^2 = \Theta = 0$, and hence $l^2 - (l_z - aE)^2 = 0$. This motivates the following definition:
\begin{equation}
    \sin \sigma = \frac{l_z - aE}{l},
\end{equation}
and thus $d l_z = \sqrt{l^2 - (l_z - a E)^2} d \sigma$. The momentum integration element reads
\begin{equation}
    dp_t dp_r dp_\theta dp_\varphi = \frac{\rho^2 \sqrt{l^2 - (l_z - a E)^2}}{\sqrt{R}\sqrt{\Theta}} (m dm) d E (l dl) d \sigma.
\end{equation}
At the equatorial plane we get simply
\begin{equation}
    \frac{\rho^2 \sqrt{l^2 - (l_z - a E)^2}}{\sqrt{R}\sqrt{\Theta}} = \frac{r^2}{\sqrt{R}}.
\end{equation}
Finally,
\begin{equation}
    \mathrm{dvol}_x(p) = \frac{1}{\sqrt{R(r)}} (m dm) d E (l dl) d \sigma,
\end{equation}
where $\sigma = \pm \pi/2$ and we have used the fact that $\sqrt{-\mathrm{det} (g^{\mu \nu})} = 1/r^2$ and $\sqrt{\Theta} = \sqrt{l^2 - (l_z - aE)^2}$ at the equatorial plane. In order to control the sign of $\sigma$ at the equatorial plane we assume a convention with $l \ge 0$ and $l = \epsilon_\sigma (l_z - a E)$ or, equivalently, $l_z = \epsilon_\sigma l + a E$. Thus $\sigma = \epsilon_\sigma \pi/2$. This yields, at the equatorial plane,
\begin{eqnarray}
    R(r) & = & (r^2 E - \epsilon_\sigma a l)^2 - \Delta (m^2 r^2 + l^2) \nonumber \\
    & = & M^4 m^2 \tilde R(\xi),
\end{eqnarray}
where $\tilde R(\xi)$ is given by Eq.\ (\ref{Rtilde}).

Using dimensionless variables (\ref{dimensionless}) we get, at the equatorial plane,
\begin{equation}
    \mathrm{dvol}_x(p) = \frac{m^3 \lambda}{\sqrt{\tilde R(\xi)}} dm d\varepsilon d\lambda d\sigma.
\end{equation}

In order to compute integrals (\ref{Jmu4d}) and (\ref{Tmunu4d}), one needs to invert relations (\ref{pmutoconst}). For the motion confined to the equatorial plane outside of the black hole horizon, one gets
\begin{eqnarray}
p_t & = & - E = - m \varepsilon, \\
p_r & = & \epsilon_r \frac{\sqrt{R}}{\Delta} = \frac{\epsilon_r m \sqrt{\tilde R}}{\xi^2 - 2 \xi + \alpha^2}, \\
p_\theta & = & 0, \\
p_\varphi & = & l_z = \epsilon_\sigma l + a E = M m (\epsilon_\sigma \lambda + \alpha \varepsilon),
\end{eqnarray}
where we have expressed the results in terms of $m$, $\varepsilon$, and $\lambda$. It is also convenient to have an explicit expression for $p^r$, which reads
\begin{equation}
    p^r = \frac{\epsilon_r m \sqrt{\tilde R}}{\xi^2}.
\end{equation}

We now take
\begin{eqnarray}
    F & = & A \delta(m - m_0) \frac{\xi}{m \lambda} \left[\delta(\sigma - \pi/2) + \delta(\sigma + \pi/2)\right] \nonumber \\
    && \times \exp(-\beta \varepsilon)
\end{eqnarray}
and
\begin{eqnarray}
    \mathcal F & = & \frac{A}{m_0 M \lambda} \delta(m - m_0) \left[\delta(\sigma - \pi/2) + \delta(\sigma + \pi/2)\right] \nonumber \\
    && \times \delta \left( \theta - \frac{\pi}{2} \right) \exp(-\beta \varepsilon).
\label{F4d}
\end{eqnarray}
Note that at the equatorial plane
\begin{equation}
\label{deltasigma}
    \frac{\xi}{m \lambda} \left[\delta(\sigma - \pi/2) + \delta(\sigma + \pi/2)\right] = \delta(p_z).
\end{equation}
This relation follows from equations $p_\theta = - r p_z$ and $l^2 = p_\theta^2 + (p_\varphi + a p_t )^2$, which are satisfied at the equatorial plane. Thus
\begin{equation}
    p_z = - \epsilon_\theta \frac{l \cos \sigma}{r},
\end{equation}
where $\epsilon_\theta$ is the sign of $p_\theta$. One can check that $\mathcal F$ given by Eq.\ (\ref{F4d}) satisfies Eq.\ (\ref{vlasoveq4d}).

\begin{widetext}
This gives
\begin{eqnarray}
\label{jt}
J_t(\xi) & = & - \sum_{\epsilon_\sigma = \pm 1} A m_0^3 \xi \int \exp(-\beta \varepsilon) \frac{\varepsilon}{\sqrt{\tilde R(\xi)}} d\varepsilon d\lambda, \\
J_r(\xi) & = & \sum_{\epsilon_\sigma = \pm 1} \frac{A m_0^3 \xi}{\xi^2 - 2 \xi + \alpha^2} \int \epsilon_r \exp(- \beta \varepsilon) d \varepsilon d \lambda, \\
J_\theta(\xi) & = & 0, \\
J_\varphi(\xi) & = & \sum_{\epsilon_\sigma = \pm 1} A M m_0^3 \xi \int \exp( - \beta \varepsilon ) \frac{\epsilon_\sigma \lambda + \alpha \varepsilon}{\sqrt{\tilde R(\xi)}} d\varepsilon d\lambda,
\end{eqnarray}
in agreement with Eqs.\ (\ref{Jmugeneral3d}).
\end{widetext}

In the same four dimensional setup we can also recover the formulas for the accretion rates. A unit vector normal to the boundary $r = r_1$ and tangent to the hypersurface of constant time has the components $\hat n_\mu = (0, \pm \varrho/\sqrt{\Delta},0,0)$. The determinant of the metric induced at $r = r_1$ reads $\mathrm{det} \, \hat \gamma = \Delta \varrho^2 \sin^2 \theta$.

The flux of the vector field $\hat V^\mu$ through the boundary at $r = r_1$ can be written as
\begin{equation}
    \int_{r = r_1} d^3 y \sqrt{|\mathrm{det} \,\gamma|} \hat n_\mu \hat V^\mu = \pm \int_{r = r_1} dt d\theta d\varphi \varrho^2 \sin \theta \hat V^r,
\end{equation}
where, as in our previous calculation, we choose the minus sign. This gives rise to the accretion rate
\begin{equation}
    \dot V = - \int_{r = r_1} d \theta d \varphi \varrho^2 \sin \theta \hat V^r.
\end{equation}
For the current $\hat V^\mu$ confined to the equatorial plane with $\hat V^r = V^r \delta(z) = V^r \delta (\theta - \pi/2)/r$, where $V^r$ does not depend on $\theta$, we get
\begin{equation}
    \dot V = - r_1 \int_{r = r_1} d \varphi V^r.
\end{equation}
The remaining part of the derivation follows the footsteps of our 3 dimensional calculation. The only difference is that in the 4 dimensional case, the conserved vector currents are given by $\mathcal{J}^\mu$, $\mathcal{T}\indices{^\mu_\nu} \xi^\nu$, and  $\mathcal{T}\indices{^\mu_\nu} \chi^\nu$, where the Killing vectors associated with the four dimensional metric $g$ have the components $\xi^\mu = (\xi^t,\xi^r,\xi^\theta,\xi^\varphi) = (1,0,0,0)$ and $\chi^\mu = (\chi^t,\chi^r,\chi^\theta,\chi^\varphi) = (0,0,0,1)$. This again leads to accretion rates defined as in Eqs.\ (\ref{dotM_nonparam}), (\ref{dotE_nonparam}), and (\ref{dotL_nonparam}).

\section{Maxwell-J\"{u}ttner distribution for the gas confined to a single plane in the Minkowski spacetime}
\label{sec:minkowski}

Thermodynamic relations characterizing the Maxwell-J\"{u}ttner distribution in arbitrary dimensions are given, e.g., in \cite{Acuna}, Sec.\ 3.3. In this appendix, we derive some of these relations for 2 + 1 dimensions, and we do so mainly to keep the same conventions as in the remainder of this paper. In particular, we compute the rest-mass surface density of a gas in thermal equilibrium, confined to a single plane in the Minkowski spacetime, as well as the corresponding energy surface density.

We work in Cartesian coordinates $(t,x,y,z)$. The Minkowski metric has the form
\begin{equation}
    g = -dt^2 + dx^2 + dy^2 + dz^2.
\end{equation}
We assume the same notation as in Appendix \ref{appendixA}. The particle current density is denoted by $\mathcal{J}^\mu$, and the energy-momentum tensor by $\mathcal{T}^{\mu\nu}$. They are defined by Eqs.\ (\ref{Jmufull}) and (\ref{Tmunufull}) with $\mathrm{dvol}_x(p) = dp_t dp_x dp_y dp_z$. Surface quantities $J^\mu$ and $T^{\mu\nu}$ are defined as in Eqs.\ (\ref{surfacequant}). Distribution functions $\mathcal F$ and $F$ are related by Eq.\ (\ref{twodf}).

The Maxwell-J\"{u}ttner distribution of the gas confined to a single plane $z = 0$ is defined by the distribution function
\begin{eqnarray}
    F & = & A \delta(\sqrt{-p_\mu p^\mu} - m_0) \delta(p_z) \exp \left( \frac{\beta}{m_0} p_t \right) \nonumber \\
    & = & A \delta \left( \sqrt{p_t^2 - \mathbf p^2} - m_0\right) \delta(p_z) \exp \left( \frac{\beta}{m_0} p_t \right).
    \label{Fplane}
\end{eqnarray}

The rest-mass density is equal to $- m_0 \mathcal{J}_t$. The rest-mass surface density, used in this paper, reads $\rho_s = - m_0 J_t$. Similarily, the energy density equals $- \mathcal{T}\indices{^t_t}$, and the energy surface density is given by $\varepsilon_s = - T\indices{^t_t}$.

The component $J_t$ can be computed as
\begin{eqnarray}
    J_t & = & A \int  \delta \left( \sqrt{p_t^2 - \mathbf p^2} - m_0\right) \delta(p_z) \nonumber \\
    && \times \exp \left( \frac{\beta}{m_0} p_t \right) p_t dp_t dp_x dp_y dp_z,
\end{eqnarray}
where $\mathbf p^2 = p_x^2 + p_y^2 + p_z^2$, and where we only take into account future-pointing momenta. In the first step we use the identity
\begin{equation}
    \delta \left( \sqrt{p_t^2 - \mathbf p^2} - m_0 \right) = \frac{m_0}{\sqrt{m_0^2 + \mathbf p^2}} \delta \left( p_t + \sqrt{m_0^2 + \mathbf p^2} \right).
\end{equation}
Integrating over $p_t$, we obtain
\begin{equation}
    J_t = - A m_0 \int \delta(p_z) \exp \left( -\frac{\beta}{m_0} \sqrt{m_0^2 + \mathbf p^2} \right) d p_x d p_y d p_z.
\end{equation}
Next, a straightforward integration over $p_z$ gives
\begin{equation}
    J_t = - A m_0 \int \exp \left( -\frac{\beta}{m_0} \sqrt{m_0^2 + p_x^2 + p_y^2} \right) d p_x d p_y.
\end{equation}
The next step is standard---we change momentum coordinates $(p_x,p_y)$ to polar ones $(\zeta,\vartheta)$, defined by $p_x = \zeta \cos \vartheta$, $p_y = \zeta \sin \vartheta$. This gives $dp_x dp_y = \zeta d \vartheta d \zeta$, where $p_x^2 + p_y^2 = \zeta^2$. Hence
\begin{equation}
    J_t = - 2 \pi A m_0 \int_0^\infty \exp \left( -\frac{\beta}{m_0} \sqrt{m_0^2 + \zeta^2} \right) \zeta d\zeta.
\end{equation}
We can now substitute $\chi = \frac{\beta}{m_0} \sqrt{m_0^2 + \zeta^2}$. Thus, $\zeta d\zeta = (m_0/\beta)^2 \chi d \chi$, and
\begin{eqnarray}
    J_t & = & - \frac{2 \pi A m_0^3}{\beta^2} \int_\beta^\infty \exp \left( - \chi \right) \chi d\chi \nonumber \\
    & = & - 2 \pi A m_0^3 \frac{1 + \beta}{\beta^2} \exp(-\beta).
\label{jtflat}
\end{eqnarray}
The particle number surface density reads
\begin{equation}
    n_s = 2 \pi A m_0^3 \frac{1 + \beta}{\beta^2} \exp(-\beta),
\end{equation}
and the rest-mass surface density can be expressed as $\rho_s = m_0 n_s$.

In a similar fashion, one can compute the expression for the energy surface density
\begin{equation}
   \varepsilon_s =  - T\indices{^t_t} = - \int F(p_\nu) p^t p_t dp_t dp_x dp_y dp_z,
\end{equation}
where $F(p_\nu)$ is given by Eq.\ (\ref{Fplane}). The result reads
\begin{equation}
    \varepsilon_s = 2 \pi A m_0^4 \frac{2 + 2 \beta + \beta^2}{\beta^3} \exp(- \beta).
\end{equation}

It is also worth noticing that the equivalent of the pressure, i.e., $P_s = T\indices{^x_x} = T\indices{^y_y}$ reads
\begin{eqnarray}
    P_s & = & \int F(p_\nu) p^x p_x dp_t dp_x dp_y dp_z\\
    & = & 2 \pi A m_0^4 \frac{1+\beta}{\beta^3} \exp(-\beta).
\end{eqnarray}
Thus
\begin{equation}
    \frac{P_s}{n_s} = \frac{m_0}{\beta} = k_\mathrm{B} T.
\end{equation}
We see that the two-diemensional Maxwell-J\"{u}ttner distribution gives rise to the standard ideal gas equation.

The specific enthalpy associated with the two dimensional Maxwell-J\"{u}ttner distribution is
\begin{equation}
    \frac{\varepsilon_s + P_s}{\rho_s} = \frac{3 + 3 \beta + \beta^2}{\beta (1 + \beta)}.
\end{equation}

These results should be contrasted with the standard formula obtained for the the gas not restricted to one plane, but filling the whole space uniformly. In the latter case, we assume
\begin{equation}
    \mathcal F = A^\prime \delta \left( \sqrt{p_t^2 - \mathbf p^2} - m_0\right) \exp \left( \frac{\beta}{m_0} p_t \right).
\end{equation}
This gives, as before
\begin{equation}
     \mathcal J_t = - A^\prime m_0 \int \exp \left( -\frac{\beta}{m_0} \sqrt{m_0^2 + \mathbf p^2} \right) d p_x d p_y d p_z.
\end{equation}
Changing to spherical momentum coordinates we get
\begin{equation}
    \mathcal J_t = - 4 \pi A^\prime m_0 \int_0^\infty \exp \left( -\frac{\beta}{m_0} \sqrt{m_0^2 + \zeta^2} \right) \zeta^2 d \zeta.
\end{equation}
The substitution $\chi = \frac{\beta}{m_0} \sqrt{m_0^2 + \zeta^2}$ now gives
\begin{eqnarray}
    \mathcal J_t & = & - \frac{4 \pi A^\prime m_0^4}{\beta^2} \int_\beta^\infty \exp (-\chi) \chi \sqrt{\left( \frac{\chi}{\beta} \right)^2 - 1} d \chi \nonumber \\
    & = & - 4 \pi A^\prime m_0^4 \frac{K_2(\beta)}{\beta}.
\end{eqnarray}
Thus, the particle number density for the three dimensional Maxwell-J\"{u}ttner distribution can be written as
\begin{equation}
\label{3dn}
    n = 4 \pi A^\prime m_0^4 \frac{K_2(\beta)}{\beta}.
\end{equation}

\end{document}